\documentclass[fleqn,usenatbib]{mnras}
\usepackage{newtxtext,newtxmath, longtable}

\usepackage[T1]{fontenc}

\DeclareRobustCommand{\VAN}[3]{#2}
\let\VANthebibliography\thebibliography
\def\thebibliography{\DeclareRobustCommand{\VAN}[3]{##3}\VANthebibliography}

\usepackage{graphicx}	
\usepackage{amsmath}	
 \usepackage{longtable}
 \usepackage{xtab,afterpage}

\DeclareUnicodeCharacter{2212}{-}

\title[Finding Radio Transients with Citizen Science]{Bursts from Space: MeerKAT - The first citizen science project dedicated to commensal radio transients}
\author[A. Andersson et al.]{Alex Andersson,$^{1}$\thanks{E-mail: alexander.andersson@physics.ox.ac.uk}
Chris Lintott,$^{1}$
Rob Fender,$^{1,2}$
Joe Bright,$^{1}$
Francesco Carotenuto,$^{1}$
Laura Driessen,$^{3,4,5}$
\newauthor{Mathilde Espinasse,$^{6}$
Kelebogile Gaseahalwe,$^{2,7}$
Ian Heywood,$^{1,8,9}$
Alexander J. van der Horst,$^{10,11}$
Sara Motta,$^{12}$
}
\newauthor{Lauren Rhodes,$^{1}$
Evangelia Tremou ,$^{13}$
David R. A. Williams,$^{4}$
Patrick Woudt,$^{2}$
Xian Zhang,$^{14,15}$
}
\newauthor{Steven Bloemen,$^{16}$
Paul Groot,$^{2,7,16}$
Paul Vreeswijk,$^{16}$
Stefano Giarratana,$^{17,18}$
Payaswini Saikia,$^{19}$}
\newauthor{Jonas Andersson,$^{20}$
Lizzeth Ruiz Arroyo,$^{20}$
Loïc Baert,$^{20}$
Matthew Baumann,$^{20}$
Wilfried Domainko,$^{20}$}
\newauthor{Thorsten Eschweiler,$^{20}$
Tim Forsythe,$^{20}$
Sauro Gaudenzi,$^{20}$
Rachel Ann Grenier,$^{20}$
Davide Iannone,$^{20}$}
\newauthor{Karla Lahoz,$^{20}$
Kyle J. Melville,$^{20}$
Marianne De Sousa Nascimento,$^{20}$
Leticia Navarro,$^{20}$
Sai Parthasarathi,$^{20}$}
\newauthor{Piilonen,$^{20}$
Najma Rahman,$^{20}$
Jeffrey Smith,$^{20}$
B. Stewart,$^{20}$
Newton Temoke,$^{20}$
Chloe Tworek$^{20}$ and
}
\newauthor{Isabelle Whittle$^{20}$}
\\
$^{1}$Astrophysics, Department of Physics, University of Oxford, Denys Wilkinson Building, Keble Road, Oxford OX1 3RH, UK\\
$^{2}$Department of Astronomy, University of Cape Town, Private Bag X3, Rondebosch 7701, South Africa\\
$^{3}$CSIRO, Space and Astronomy, PO Box 1130, Bentley, WA 6102, Australia\\
$^{4}$Jodrell Bank Centre for Astrophysics, Department of Physics and Astronomy, The University of Manchester, Manchester, M13 9PL, UK\\
$^{5}$Sydney Institute for Astronomy, School of Physics, The University of Sydney, NSW 2006, Australia\\
$^{6}$AIM/CEA Paris-Saclay, Universite Paris Diderot, CNRS, F-91191 Gif-sur-Yvette, France\\
$^{7}$South African Astronomical Observatory, PO Box 9, Observatory 7935, South Africa\\
$^{8}$Department of Physics and Electronics, Rhodes University, PO Box 94, Makhanda 6140, South Africa\\
$^{9}$South African Radio Astronomy Observatory, 2 Fir Street, Observatory 7925, South Africa\\
$^{10}$Department of Physics, The George Washington University, 725 21st Street NW, Washington, DC 20052, USA\\
$^{11}$Astronomy, Physics and Statistics Institute of Sciences (APSIS), 725 21st Street NW, Washington, DC 20052, USA\\
$^{12}$Istituto Nazionale di Astrofisica, Osservatorio Astronomico di Brera, via E. Bianchi 46, 23807 Merate (LC), Italy\\
$^{13}$National Radio Astronomy Observatory, Socorro, NM 87801, USA\\
$^{14}$Shanghai Astronomical Observatory, Chinese Academy of Sciences, 80 Nandan Road, Shanghai 200030, China\\
$^{15}$University of Chinese Academy of Sciences, 19A Yuquanlu, Beijing 100049, China\\
$^{16}$Department of Astrophysics/IMAPP, Radboud University, P.O. 9010, 6500 GL, Nĳmegen, The Netherlands\\
$^{17}$Dipartimento di Fisica e Astronomia, Università degli Studi di Bologna, Via Gobetti 93/2, 40129 Bologna, Italy\\
$^{18}$INAF – Istituto di Radioastronomia, Via Gobetti 101, 40129 Bologna, Italy\\
$^{19}$Center for Astro, Particle and Planetary Physics, New York University Abu Dhabi, PO Box 129188, Abu Dhabi, UAE\\
$^{20}$Citizen Scientist, Zooniverse c/o University of Oxford, Keble Road, Oxford OX1 3RH, UK\\
}

\date{Accepted XXX. Received YYY; in original form ZZZ}

\pubyear{2022}

\begin{document}
\label{firstpage}
\pagerange{\pageref{firstpage}--\pageref{lastpage}} 
\maketitle

\begin{abstract}
The newest generation of radio telescopes are able to survey large areas with high sensitivity and cadence, producing data volumes that require new methods to better understand the transient sky. Here we describe the results from the first citizen science project dedicated to commensal radio transients, using data from the MeerKAT telescope with weekly cadence. \textit{Bursts from Space: MeerKAT} was launched late in 2021 and received $\sim$89000 classifications from over 1000 volunteers in 3 months. Our volunteers discovered 142 new variable sources which, along with the known transients in our fields, allowed us to estimate that at least 2.1 per cent of radio sources are varying at 1.28 GHz at the sampled cadence and sensitivity, in line with previous work. We provide the full catalogue of these sources, the largest of candidate radio variables to date. Transient sources found with archival counterparts include a pulsar (B1845-01) and an OH maser star (OH 30.1–0.7), in addition to the recovery of known stellar flares and X-ray binary jets in our observations. Data from the MeerLICHT optical telescope, along with estimates of long time-scale variability induced by scintillation, imply that the majority of the new variables are active galactic nuclei. This tells us that citizen scientists can discover phenomena varying on time-scales from weeks to several years. The success both in terms of volunteer engagement and scientific merit warrants the continued development of the project, whilst we use the classifications from volunteers to develop machine learning techniques for finding transients.

\end{abstract}

\begin{keywords}
radio continuum: transients -- radio continuum: galaxies -- radio continuum: general -- surveys
\end{keywords}



\section{Introduction}

The latest generation of radio telescopes provide us with unprecedented detail of the radio sky. Regular, wide-field images from highly sensitive telescopes, including Square Kilometre Array (SKA) pathfinders such as MeerKAT \citep{2016mks..confE...1J} and the Australian SKA Pathfinder \citep[ASKAP;][]{2021PASA...38....9H} allow us to probe a wide range of physics at novel time-scales and depths. For example, stellar radio emission can provide insight into magnetic re-connection \citep{Rigney2022} and has implications for orbiting planet habitability \citep{Airapetian2017, Gunther2020}, whilst the radio afterglows of gamma-ray bursts constrain jet physics and kinetic feedback of the most violent eruptions in the Universe \citep[e.g.][]{Rhodes2021}. Similarly, fast radio bursts can probe the baryonic content of the Universe \citep{2020Natur.581..391M}, whilst the afterglows of neutron star mergers can provide key insights into the structure of relativistic jets \citep{2021ARA&A..59..155M}. The combination of sensitivity, regular cadence and (crucially) wide field of view (FoV) allow for commensal, untargetted strategies in order to search for these known transient phenomena and new classes of objects as yet undiscovered. 

Previous investigations from both MeerKAT and other instruments have found that only a few per cent of point sources aretransient or variable above sensitivity limits at 1.4 GHz (\citealp{Ofek2011} and references therein), with source classes spanning a wide range of time-scales and physical processes. The majority of radio variables found are active galactic nuclei (AGN; see e.g. \citealp{Thyagarajan2011}), whose variations can be attributed to a combination of refractive scintillation \citep{Rickett1990} and shock-induced flaring in their jets \citep{Mooley2016}. Whilst these AGN dominate samples of variables, active or flaring stars have been found in untargetted radio surveys \citep{Mooley2016, Driessen2020, Andersson2022}, as have supernovae and GRB orphan afterglow candidates \citep{Levinson2002, 2006ApJ...639..331G}. Pulsars can vary intrinsically in the image plane - indeed some of the slowest known pulsars are discovered in imaging data \citep{2018ApJ...866...54T, Caleb2022}. Futhermore, diffractive scintillation through the interstellar medium can cause short time-scale, large amplitude variations in observations of pulsars, whilst refractive scintillation can produce lower amplitude variability occurring on time-scales of hours to years for point-like sources \citep{Rickett2001, Hancock2019}. There are also numerous accounts of radio transients being discovered without clear progenitor systems or multiwavelength counterparts \citep{Bower2007, Stewart2016, Murphy2017}. These include the elusive sources near the Galactic centre \citep{1976Natur.261..476D, Zhao1992,Hyman2005, Chiti2016} - including the newest such transient found by ASKAP \citep{Wang2021}. The serendipitous discoveries, elusive nature and broad physics at play in this zoo of radio transients all point towards the need for new searches and the development of novel methods to maximise the science yield of our observations.

\defcitealias{Driessen2022}{D22}
\defcitealias{Rowlinson2022}{R22}

ThunderKAT\footnote{\textbf{T}he \textbf{Hun}t for \textbf{D}ynamic and \textbf{E}xplosive \textbf{R}adio transients with Meer\textbf{KAT}} \citep{Fender2016} is a large survey project dedicated to monitoring radio transients with MeerKAT. The ThunderKAT team regularly observes known transients such as X-ray binaries, cataclysmic variables and gamma-ray bursts (XRBs, CVs and GRBs respectively). The large field of view of MeerKAT ($\sim1$ square degree at 1.28 GHz), sampled at approximately weekly cadences with high sensitivity also allows for unprecedented commensal searches for transients, variables and other ancilliary science. \cite{Driessen2020}, \cite{Driessen2021} and \cite{Andersson2022} describe the first commensal transients found with MeerKAT, detailing flaring and quiescent behaviour from stellar systems. Similarly, \cite{Driessen2022} and \citet[][ hereafter D22 and R22 respectively]{Rowlinson2022} make use of the best-sampled ThunderKAT fields surrounding XRBs GX339$-4$ and \textit{MAXI} J1820+070 to discover new radio variables including pulsars and variable AGN. Images of short GRB fields have been searched for transients at both fast and slower time-scales in Chastain et al. (submitted), wherein there are many newly described variables, of which most are likely scintillating AGN.  The ThunderKAT survey also makes use of MeerLICHT \citep{Bloemen2016}, a robotic facility (65cm primary mirror) whose goals include shadowing MeerKAT observations, providing spatial and temporal coverage of the optical sky to complement the radio data.

Despite these methods of searching for radio transients bearing fruit, they are not optimal. Firstly, the volume of data to analyse is far greater than any one person can achieve by eye on reasonable time-scales. In the 4 years since operations began, ThunderKAT has observed over 30 XRBs in total at weekly cadence, typically following each source for over a month. This results in over 100 TB of raw data to reduce and analyse, producing over 500 final images. Each image then contains of order several hundred sources, from just a single 15 minute observation. ThunderKAT also has memoranda of understanding with many of the other Large Survey Projects on MeerKAT, such as LADUMA \citep{2016mks..confE...4B}, MIGHTEE \citep{2016mks..confE...6J} and MHONGOOSE \citep{2016mks..confE...7D}, to use their data commensally. As a result there are many hundreds of observations in the growing archive, in which transients may reside, probing right down to 1$\sigma$ sensitivity limits $\sim1\mu$Jy \citep{2022MNRAS.509.2150H}.  These data overload issues are only exacerbated when imaging on shorter time-scales, including  down to the 8s integration time of MeerKAT, as is currently being tested within ThunderKAT (e.g. \citealp{Caleb2022}; Chastain et al. submitted; Fijma et al. in prep). 

Radio observations are not free from false positives. Two main causes of these false positives are the non-Gaussian artefacts that typically occur around bright sources in radio images, and the changes in the point spread function (PSF, or restoring beam) caused by differing elevations over a set of observations, which induces non-intrinsic variability in the measurements of resolved objects. As these issues might plague only one observation in a dataset, they can lead to measurements easily confused for bona fide transients by automated methods.

One method to search for radio transients is by harnessing the power of citizen science. Citizen science projects hosted on the Zooniverse\footnote{\url{zooniverse.org}} have been highly successful in transient astrophysics and astronomy more generally, starting with the original \textit{Galaxy Zoo} \citep{Lintott2008}. Since then the hundreds of public projects have received over 700 million classifications from 2.5 million users (taken from the website's live tracker). In transient astronomy specifically, \cite{Wright2017}'s \textit{Supernova Hunters} combined a neural network with human classifications to outperform either classifier alone and is still discovering supernovae, over six years since launch\footnote{see \url{https://www.wis-tns.org/object/2022aeee}}. Similarly, \textit{Citizen ASAS-SN} users have discovered $>$10,000 new variable sources that are not present in the existing star catalogs of the southern hemisphere \citep{Christy2021}. The Zooniverse's Talk feature (project specific forums where users, moderators and experts can discuss individual subjects, classifications and so on) provides room for novel discovery space - classic \textit{Galaxy Zoo} examples include the `Green Peas', a class of compact galaxies with extremely high star formation rates \citep{Cardamone2009} and  `Hanny's Voorwerp', an extended region of gas ionised by the now-faded AGN of IC 2497 \citep{Lintott2009}. Similar finds from other projects include unusual variable stars and new classes of systematic noise in LIGO/Virgo detectors \citep{Christy2021, Zevin2017}. In this work we will describe the first citizen science project dedicated to radio transients. The aims of this project are to discover new transients, eliminate spurious false positives and provide complementary analysis to other commensal search methods, as well as allowing us to assess the viability of further citizen science work.

In section \ref{sec:TraP} we detail the observations and processing prior to the Zooniverse project launch discussed in section \ref{sec:citsci}, the results of which are found in section \ref{sec:results}. We search for counterparts to radio variables in section \ref{sec:mw} before the discussion and conclusions found in sections \ref{sec:disc} and \ref{sec:conclude}.

\section{ThunderKAT Observations and Pre-processing}
\label{sec:TraP}

Our observations consist of a subset of ThunderKAT XRB images, based on which datasets were available at the time of research and contained more than a few epochs. The observations used in this work were taken between mid-2018 and late 2021. Generally, the observing strategy is determined by reports from X-ray facilities of activity from an XRB, which is then observed at weekly cadence by ThunderKAT in 15 minute images. Each observing block consists of first scanning a primary calibrator, then bookending source observations with phase calibrator observations.  A table of the 11 datasets used in this study, with the number of observations in each, is given in Table \ref{table:obs}. The varying number of epochs between datasets is a direct result of the radio activity of the central XRB i.e. if a source is seen to fade below detection it is removed from the weekly scheduling block, whilst the central root mean square flux density (RMS) varies due to baseline coverage and presence of diffuse structures. The values presented are the median values of the RMS calculated across all images of a given dataset, evaluated in the central 8th of an image. It is worth mentioning that the GX339--4 field has been observed every week since ThunderKAT observations began, in contrast to the one or two outburst cycles followed for all other datasets. It is important to remember that the commensal nature of this work constrains us to whatever observational cadence was used for monitoring the XRB.

\begin{table*}
	\caption{Properties of the 11 ThunderKAT datasets used in this work. Each field's approximate Galactic latitude is given for relevance in section \ref{sec:results} and Figure \ref{fig:RISS}. The number of sources and central RMS values are calculated by the \textsc{TraP} (see section \ref{sec:TraP}).}
	\label{table:obs}
\begin{tabular}{cccccccc}
\hline
Dataset/Central XRB & Galactic        & Epochs & Duration                & Number of  & Average Central  & XRB Paper  \\
& Latitude$ (\degr)$&&& \textsc{TraP} Sources &  RMS ($\mu$Jy) & \\

\hline
GX339--4   & --04.33      & 167    & 2018-04-14 -- 2021-10-31 & 714                    & 35                         & \cite{Tremou2020};     \\
&&&&&& Tremou et al. in prep \\
\textit{MAXI} J1820+070   & +10.16   & 77     & 2018-09-28 -- 2020-11-22 & 1838                   & 26                         & \cite{Bright2020}     \\
GRS 1915+105  & --00.22    & 60     & 2018-12-08 -- 2021-04-10 & 510                    & 136                        & \cite{Motta2021}     \\
\textit{MAXI} J1348--630 & --01.10 & 51     & 2019-01-29 -- 2020-09-26 & 533                    & 45                         & \cite{Carotenuto2021} \\
\textit{MAXI} J1848--015 & --00.10 & 35     & 2021-02-28 -- 2021-11-19 & 271                    & 290                        & \cite{Tremou2021};     \\
&&&&&&Bahramian et al. submitted\\
\textit{MAXI} J1803--298 & --03.84 & 28     & 2021-05-04 -- 2021-11-19 & 1093                   & 22                         & \cite{Espinasse2021} \\
EXO1846--031     & --00.92    & 26     & 2019-08-04 -- 2020-04-10 & 366                    & 88                         & \cite{Williams2022} \\
\textit{Swift} J1858.6--0814 & --05.32    & 25     & 2018-11-10 -- 2020-03-02 & 1512                   & 22                         & \cite{Rhodes2022}     \\
4U1543--47   & +05.42    & 21     & 2021-06-19 -- 2021-11-14 & 904                    & 27                         & \cite{2021ATel14878....1Z}    \\
&&&&&& Zhang et al. in prep\\
H1743--322    & --01.83       & 11     & 2018-09-05 -- 2018-11-10 & 379                    & 52                         & \cite{Williams2020}  \\
SAX J1808.4--3658   & --08.15    & 6      & 2019-07-31 -- 2019-08-31 & 754                    & 25                         & \cite{2023MNRAS.tmp..622G}\\
\hline
\end{tabular}
\end{table*}

ThunderKAT data is typically reduced using \textsc{OxKAT} \citep{2020ascl.soft09003H}, a semi-automatic set of scripts that perform calibration, flagging and imaging of MeerKAT data. \textsc{OxKAT} makes use of several existing radio astronomy packages including \textsc{casa} \citep{McMullin2007} for tasks such as gain and bandpass calibration, self calibration and flagging, \textsc{Cubical} \citep{Kenyon2018} for further self calibration procedures, \textsc{tricolour} for further flagging \citep{2022ASPC..532..541H} and \textsc{wsclean} \citep{Offringa2014} for imaging. These steps are broken into 1st generation calibration (1GC; direction independent effects), flagging and self-calibration (2GC), with optional 3GC steps to account for direction dependent effects.
Some of the earlier observations were reduced prior to the release of \textsc{OxKAT}, however these still follow the same basic reduction of flagging using \textsc{aoflagger} \citep{Offringa2010}, bandpass, phase calibration and flux scaling in \textsc{casa}, and imaging with \textsc{wsclean}, \textsc{DDFacet} \citep{Tasse2018} or \textsc{casa}.
The commensal nature of this work means that, due to different science requirements and observational conditions for each dataset, the resultant images are heterogeneous in their properties, although mostly homogeneous within a particular field. 

\subsection{Pre-processing and subject generation}
\label{sec:TraPdetails}
Each set of images was processed using the Transients Pipeline (\textsc{TraP} \citealp{Swinbank2015}), first designed for detecting transients with LOFAR. Below is a brief description of how the pipeline works and some of the key parameters used.
The \textsc{TraP} finds sources above a user-defined threshold in a set of astronomical images, creating light curves of each unique source and calculating statistics for each source. This is done by fitting a Gaussian component to each source in every epoch and associating it with those found at that position in all previous images, updating the database as new observations are added.
Most \textsc{TraP} parameters are kept at their default values\footnote{see \url{https://tkp.readthedocs.io/en/latest/userref/config/job_params_cfg.html} for the list of all parameters.}.
The \texttt{detection\_threshold}, the signal-to-noise ratio (S/N) above which sources are detected, was fixed at 8 throughout as a trade off between detecting false positives and missing genuine sources of interest. Once a source has been found, a Gaussian component is fit from its peak down to 3$\sigma$ above the noise (\texttt{analysis\_threshold} = 3). The \texttt{expiration} i.e. the number of force fits to a position where a source was found in a previous epoch, was always kept at greater than the total number of observations in a dataset, meaning wherever a source had been found, a light curve with datapoints for all remaining time steps was created. We are interested in unresolved, point sources and their variability, so we set all source fits to be fixed at the size of the PSF via \texttt{force\_beam} = \textsc{True}. For extended regions of emission, the change in size and position angle of the PSF between observations can lead to non-intrinsic variability measurements, as discussed in section \ref{sec:citsci}. To allow for deblending we set \texttt{deblend\_nthresh} = 10, which accounts for overlaps between nearby sources such as double lobed radio galaxies. Finally, the \texttt{extraction\_radius\_pix}, describing how far `out' in the image to search for sources, was always kept to approximately 1.5$\times$ the main lobe of the primary beam, in accordance with \cite{2021ApJ...923...31S}.

Of the \textsc{TraP} outputs, the most relevant are the light curves and two statistics computed based on the time series. For a light curve consisting of $N$ data points of flux density $F_i \pm \sigma_i$, observed at frequency $\nu$, the two variability statistics are defined as

\begin{equation}
    \eta_\nu \equiv \chi^2_{N-1} =\frac{1}{N-1}\sum_{i=1}^N \frac{(F_{i,\nu} - \overline{F_\nu}^*)^2}{\sigma_i^2}
	\label{eq:eta}
\end{equation}

and

\begin{equation}
    V_\nu \equiv \frac{s_\nu}{\overline{F_\nu}} = \frac{1}{\overline{F_\nu}}\sqrt{\frac{N}{N-1}(\overline{F^2_\nu}-\overline{F_\nu}^2)}
	\label{eq:V}
\end{equation}
 
 where $s$, $\overline{F}$ and $\overline{F_\nu}^*$ denote standard deviation, mean and weighted average respectively.  Generally speaking we can use $\eta$ and $V$ to determine the statistical significance and the amplitude of variation respectively. We expect $\eta$ to correlate with average flux density; the brightest sources will have the smallest statistical uncertainties. Similarly, $V$ and $\overline{F}$ should be anti-correlated as we are only able to measure small variations for the brightest sources. We also note that flux calibration errors can lead to overinflated statistics for bright sources - e.g. systematic differences between epochs with small statistical uncertainties producing very large values of $\eta$, as equation \ref{eq:eta} does not include systematic uncertainties in its calculation.
 
 Once all observations have been processed by the \textsc{TraP}, we generate figures of the light curves and local sky around every \texttt{runningcatalog} source - that is, each of the 8874 unique entries to the database. The local sky figure is a square arcminute image centred on the weighted mean Right Ascension and Declination (RA,Dec) of each source, using the image where the source was detected at the highest S/N. Finally, these figures are uploaded to the Zooniverse platform along with some basic metadata (RA,Dec, median flux density and the time stamp of the highest S/N observation), creating the subjects for citizen scientists to classify.

\section{Citizen Science platform}
\label{sec:citsci}

\begin{figure}
	\includegraphics[width=\columnwidth]{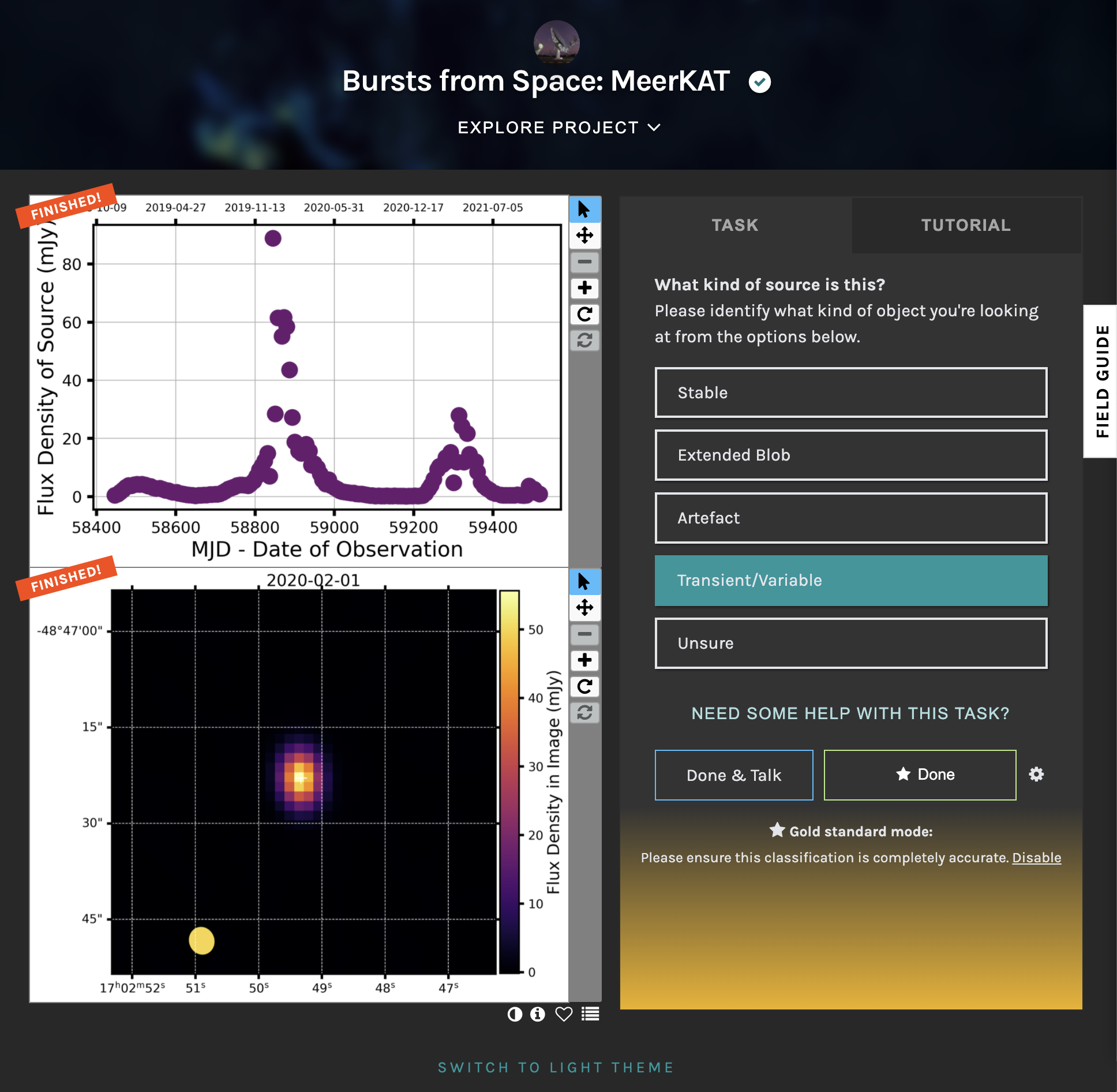}
    \caption{Classification workflow for BfS:MKT, showing the light curve and local sky figures for GX339--4.}
    \label{fig:CitSci}
\end{figure}

The \textit{Bursts from Space: MeerKAT} (hereafter BfS:MKT) Zooniverse project \footnote{\url{https://www.zooniverse.org/projects/alex-andersson/bursts-from-space-meerkat}} launched on 7th December 2021 with classifications concluding by early March 2022. During this time 1038 volunteers classified our sources using the workflow seen in Figure \ref{fig:CitSci}. Volunteers are given a tutorial to familiarise them with the data and describe the figures shown in the project, as well as accounting for common pitfalls due to figure processing - namely there are a few visualisation issues that make classifications more difficult, discussed further in section \ref{sec:disc}. There are also description pages for the project detailing the team, the telescope and the kinds of objects we are searching for so that citizen scientists can learn more about astronomy and the work we do.

\begin{figure*}
	\includegraphics[width=0.825\columnwidth]{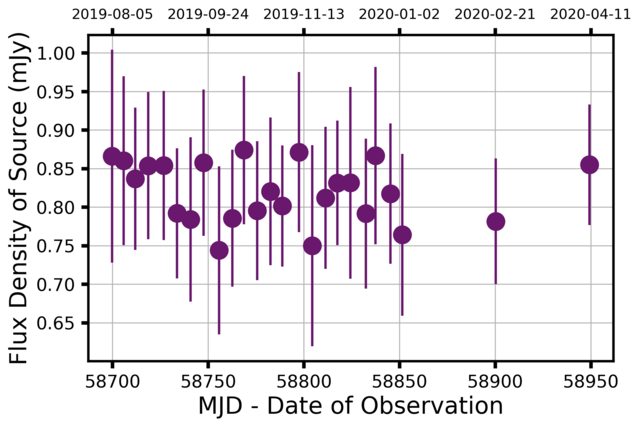} 	
         \includegraphics[width=0.825\columnwidth]{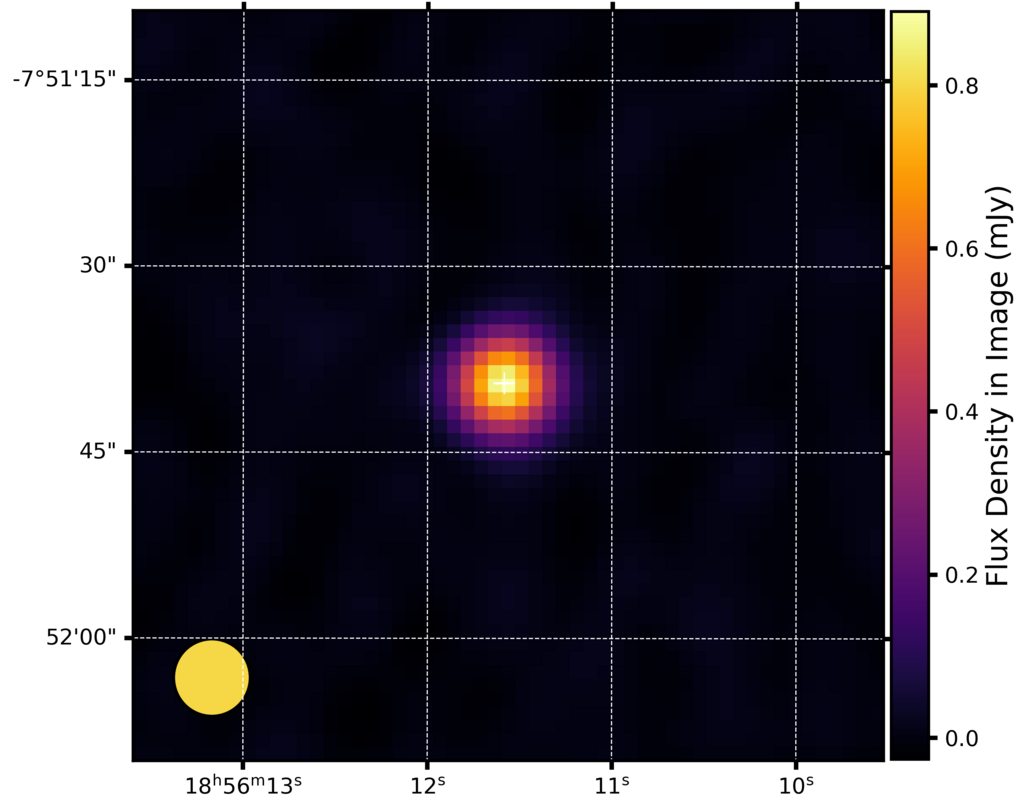}
        
        \includegraphics[width=0.825\columnwidth]{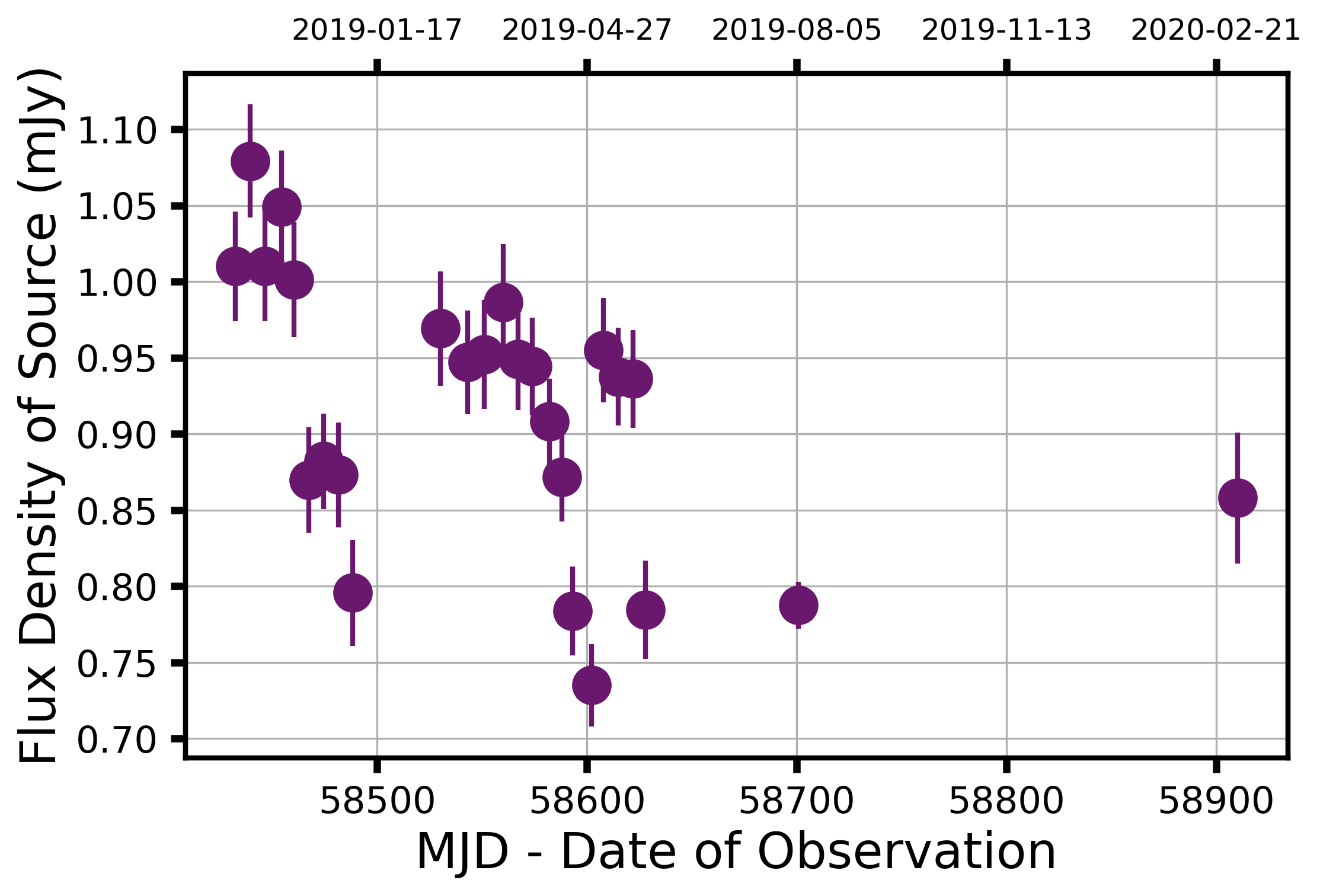}
        \includegraphics[width=0.825\columnwidth]{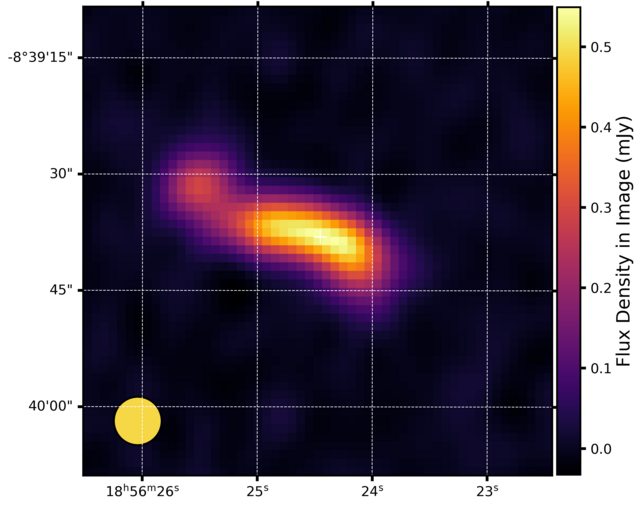}	        

        \includegraphics[width=0.825\columnwidth]{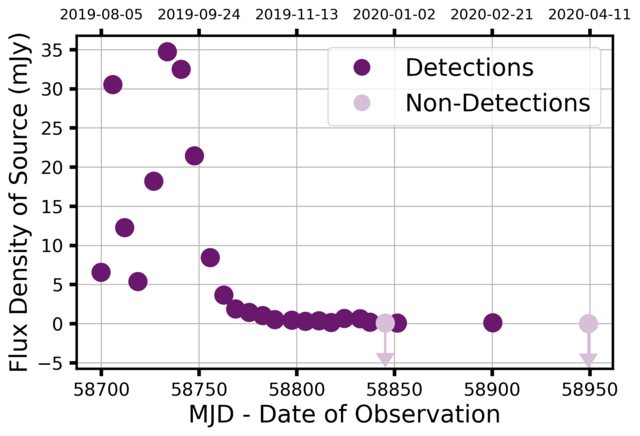}
        \includegraphics[width=0.825\columnwidth]{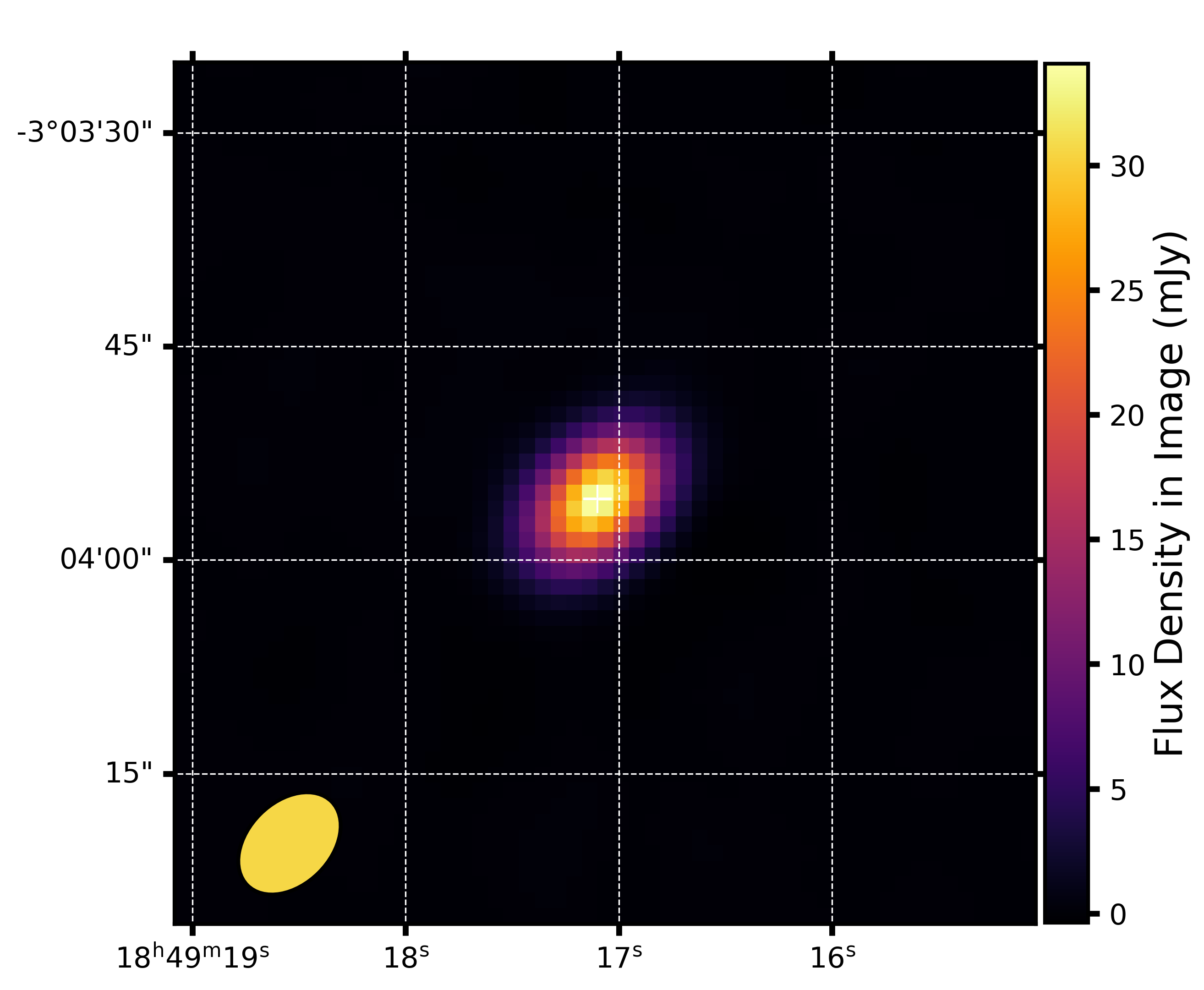}

        \includegraphics[width=0.825\columnwidth]{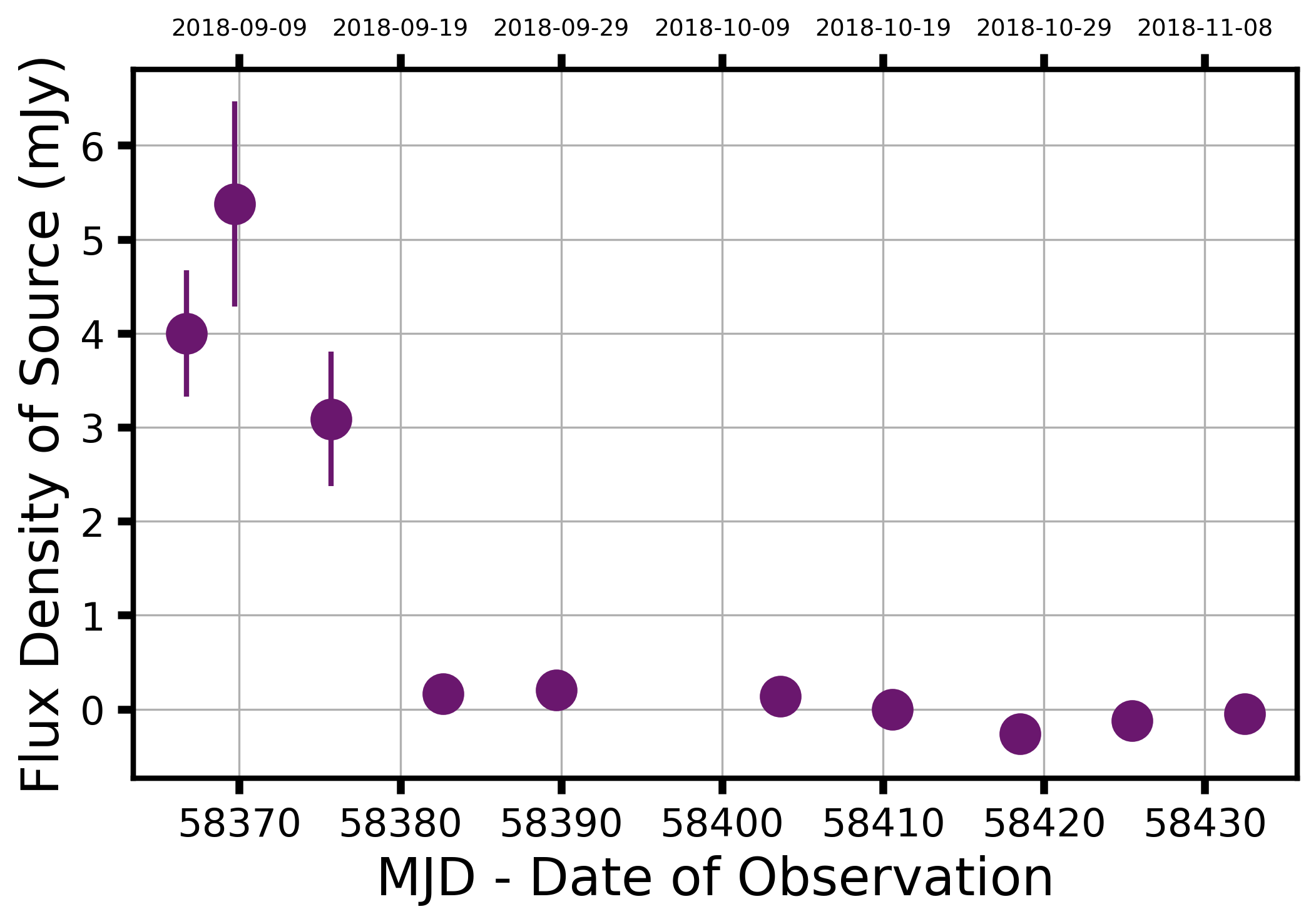}
        \includegraphics[width=0.825\columnwidth]{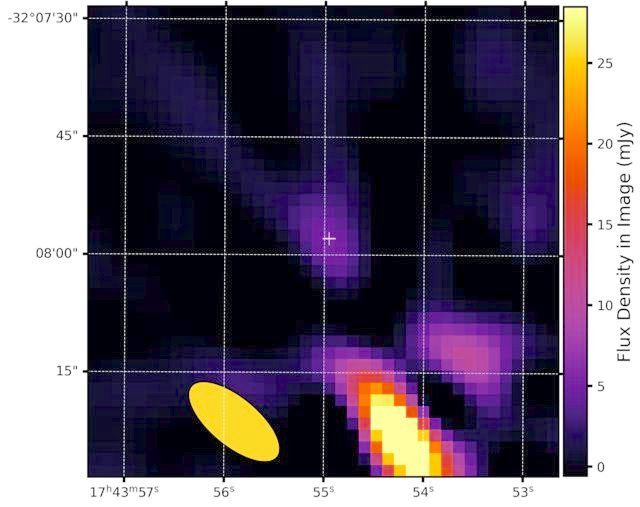}
        
    \caption{ Examples of the four observational classes within our workflow on the Zooniverse, showing both the light curve and an image of each. From top to bottom these are: \textbf{Stable} - no variation in the light curve given the error bars and a point source; \textbf{Extended} - variations caused by changes to the PSF and a source that is larger than the beam (lower left); \textbf{Transient} - a clear variable light curve for a source the same size as the PSF; \textbf{Artefact} - a spuriously transient light curve and a faint source on the outskirts of a very bright object, with non-Gaussian noise structure. The final class, \textbf{Unsure}, by definition has no archetypal characteristic so we show no figure here.}
    \label{fig:examples}
\end{figure*}

Citizen scientists were given five classes to which they can assign a source, examples of which can be seen in Figure \ref{fig:examples}. \textbf{Stable} sources are unresolved, point sources whose light curves are judged to be (within uncertainties) consistent with flat. The \textbf{Extended Blob} classification is intended to catch the resolved, extended sources, regardless of variability. We know that the changing size and angle of the PSF between epochs introduces non-intrinsic variability and so we instructed all volunteers to classify subjects they deem to be resolved as Extended Blobs - making use of the local sky figure, comparing the source size to that of the PSF in the lower left corner. The \textbf{Artefact} classification was implemented to account for any spurious, non-astrophysical sources that may be present in our images. \textbf{Transient/Variable} classifications are those we are searching for, which are point sources with variable light curves. Finally, if volunteers are uncertain if a subject fits into any of these classes - either due to visualisation issues, their own interpretation or anything other reason - they can say they are \textbf{Unsure}. We included Unsure to assess the confidence of volunteers - if a subject does not clearly fit into one class this will be seen quantitatively (not just in e.g. the Talk board). Also, without an `unsure' option, volunteers may have settled for classifying as either stable or transient, leading to an under- or over-prediction of interesting sources.
There is also a Field Guide (accessible on the main project webpage) with examples to demonstrate the type of subjects intended for each class, as well as some help text describing the rough thought process behind each source. If volunteers feel a subject is particularly interesting, or they have questions on a particular source, they can create individual threads on the dedicated Talk forum for the project, where (citizen and project) scientists can discuss.

\begin{figure}
	\includegraphics[width=\columnwidth]{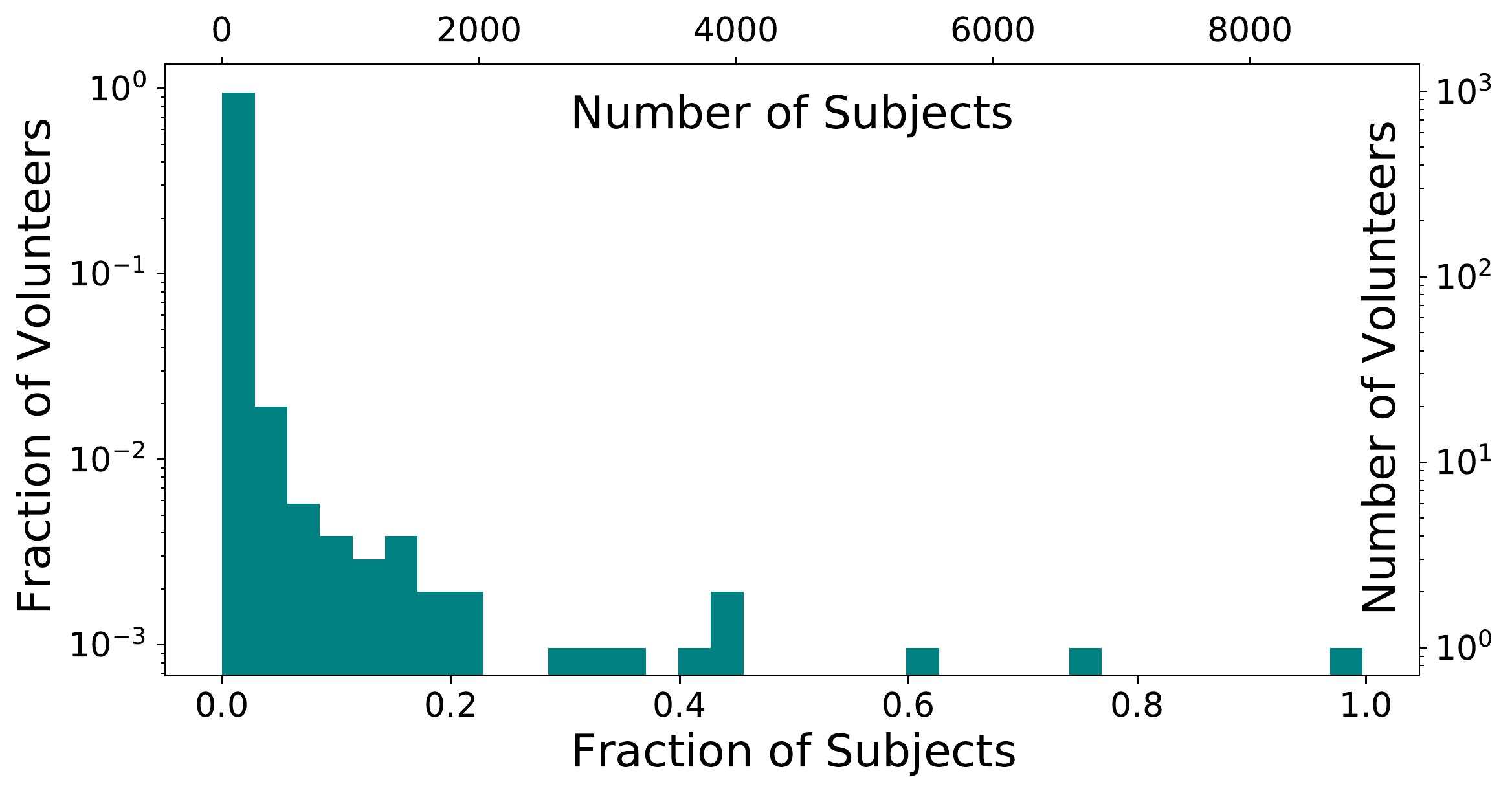}
    \caption{Histogram of the number of classifications made per user on the BfS:MKT project. As might be expected, many users perform only a few classifications, whilst a few volunteers dedicate thousands of votes to the project.}
    \label{fig:VolDist}
\end{figure}

We require 10 volunteers to classify a subject before we consider it classified, resulting in a total of 88740 classifications. These were classified over 90 days, or an average of 1 submission per 1-2 minutes. The histogram of all classifications for the project to date can be seen in Figure \ref{fig:VolDist}, showing the expected steep power law distribution of votes \citep{Spiers2019}, as well as a number of `super users' who have classified thousands of sources each. The median, mean and standard deviation of user classifications are 4.5, 86 and 490 respectively. The chosen retirement value of 10 is enough to mitigate outlier votes, but not so high that it would take many months for a single subject to be fully classified. We note here that if further iterations of the project gain as much traction as this first batch of subjects we will be able to determine what the optimal trade off could be.

 Simple aggregation is performed for this one question workflow, using the standard aggregation scripts\footnote{\url{https://aggregation-caesar.zooniverse.org/index.html}}, where we take the Boolean values for each classification and sum over every vote for each  subject. This gives us 10 votes for a given source, from which we can calculate fractional classifications and determine how many subjects are deemed to be transient/variable by some number of citizen scientists (TF = transient fraction).
 
 We set a threshold of 4/10 volunteers classifying a subject as transient/variable, reducing our sample from 8874 to 381 sources. This 0.4 threshold was chosen as a trade off between having many sources to vet and missing some low vote fraction variables. These 381 subjects were visually inspected by project scientists to confirm or reject each source as one where both volunteers and experts agree. This final vetting reduces our number of transient candidates to 168 sources (i.e. a true positive rate of 168/381 = 0.44).
 Reasons for disagreement between citizen and project scientists are numerous and include how much volunteers make use of error bars, PSFs and other parts of the figures, as well as the existence of systematic variability that is present in earlier MeerKAT observations (see e.g. the appendix of \citetalias{Driessen2022}) that would only be noticeable to experts who have compared many light curves in a given field. We will discuss what information we get from the false positives in section \ref{sec:disc}.
 
\section{Results}
\label{sec:results}

Using citizen scientists to inspect a wealth of data from the MeerKAT telescope produces 168 variable and transient radio sources which have also been vetted by project scientists. Of these 168 variables and transients found by volunteers, 142 are not detailed in previous commensal or XRB work. This constitutes one of the largest samples of radio variable candidates to date, and their positions are listed in Table \ref{tab:results}. This table provides the TF, median 1.28 GHz flux density and $\eta$ and $V$ statistics described in section \ref{sec:TraPdetails} for each of our sources, as well as the date on which they were detected at highest S/N. We strongly encourage follow-up of sources of interest to the community.

\begin{figure*}
	\includegraphics[width=\textwidth]{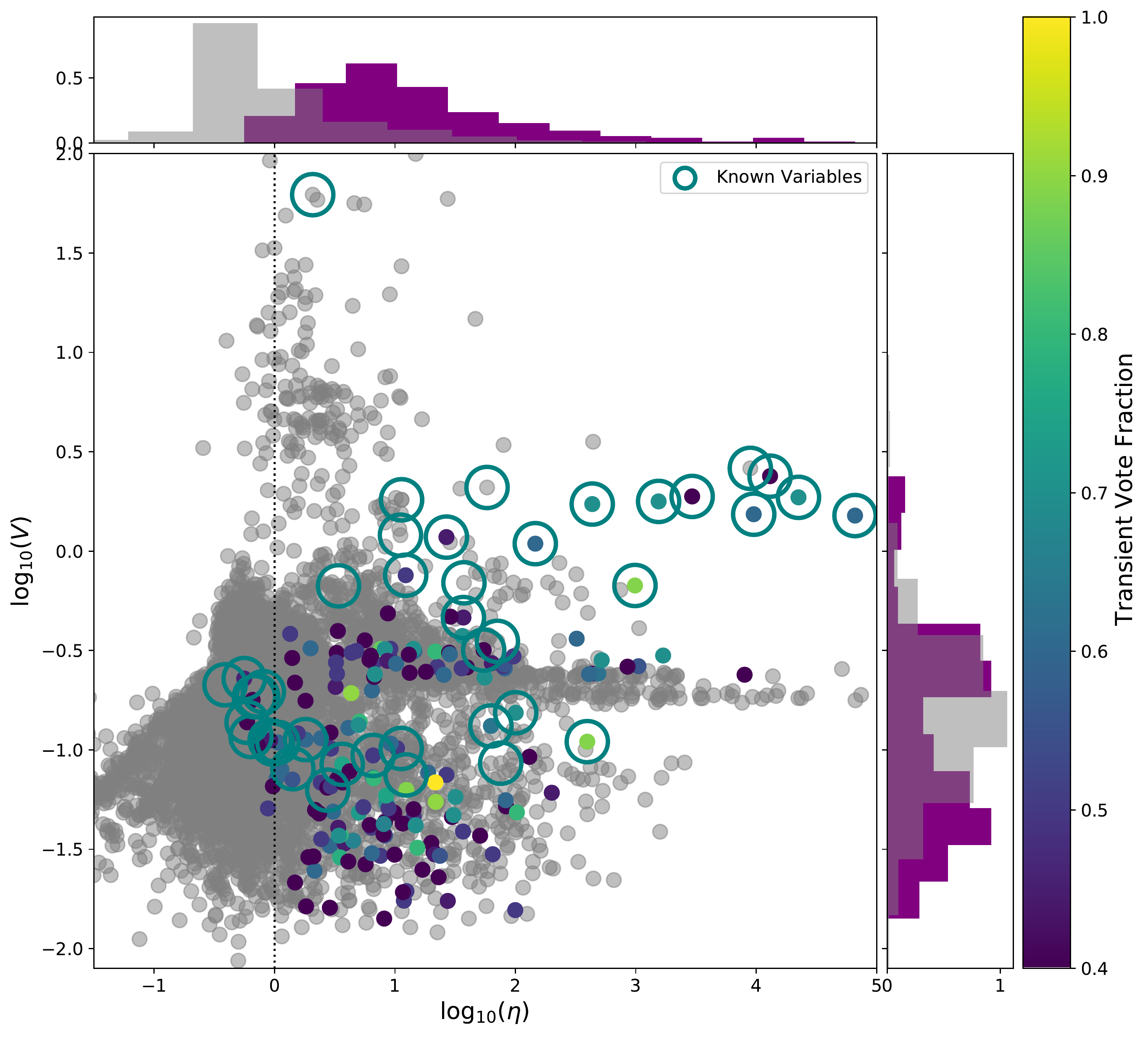}
    \caption{Variability plane for the 168 sources found by citizen scientists to be variable, along with those they find to not be so in grey. The colourbar denotes the fraction of classifications as a transient/variable source, whilst marginal distributions of $\eta$ and $V$ can also be seen. Known variable sources (e.g. XRBs) in our fields are circled. Imaging artefacts appear at low $\eta$ and high $V$, whilst flux calibration uncertainties can produce high $\eta$, low $V$ sources (due to lack of systematic uncertainty in equation \ref{eq:eta}. Most known transients are found by citizen scientists, whilst many new sources are identified and show a wide spread of values in this parameter space.}
    \label{fig:etaVPlane}
\end{figure*}

In order to characterise these sources, we can examine the variability plane defined by $\eta$ and $V$ (equations \ref{eq:eta} and \ref{eq:V}) seen in Figure \ref{fig:etaVPlane}. The structure seen in this figure consists of many sources at low significance and amplitude, some spurious artefacts at high $V$ and generally bright objects detected at high significance ($\eta$) but low variability amplitudes. Systematic differences between datasets due to their heterogeneous sampling and imaging are also present (e.g. the sources not voted as transients at log($V$) $\sim -0.6$). The most important thing to note is that there are many variable objects whose parameters are `normal' i.e. non-anomalous. This means that, were we instead to take outliers above some $N\sigma$ threshold in  $\eta$ and $V$, we could have missed sources that, upon individual inspection, appear variable or transient. So citizen scientists can find interesting variable radio sources that could have been missed by other techniques or without detailed analysis (e.g. \citetalias{Driessen2022}, \citetalias{Rowlinson2022} and Chastain et al. submitted). 
The previously known transient sources (i.e. the circled sources of Figure \ref{fig:etaVPlane}) with high $\eta$ and $V$ are almost all recovered, as are many of those found in previous studies (see sections \ref{sec:targets} and \ref{sec:commensal} for more detail), meaning that volunteers are able to recover or discover interesting sources across a range in statistical parameters. 
So our volunteers have been able to analyse large data volumes in just a few months and produce the largest sample to-date of variables from a radio telescope. 

We can use these discovered (142) and recovered (26) variables, along with the other known transients (19) in our fields, to estimate that at least 2.1 per cent of radio sources seen at 1.28 GHz with our approximately weekly observations are transient or variable. This rate is in line with previous work on radio transients; see \citet{Ofek2011} for a review, as well as \citetalias{Driessen2022} and \citet{2021ApJ...923...31S}. We also note that these different studies, including this work, probe slightly different time-scales, though all of them `long' ( $\geq$ a week). This means that our volunteers and project scientists can find comparable transient rates as previous work. This 2.1 per cent is a lower limit on the number of variables and we can estimate what fraction of variables/transient are missed by assessing which known transients are not recovered. The fraction of previously known variables not recovered is 19/45, implying the true amount of variables in our fields could be as high as $\sim5$ per cent. However, this estimate is only valid under the assumption that new variables are recovered at the same rate as the known transients in our field. In section \ref{sec:disc} we discuss the selection effects that are evident in the different kinds of light curves selected by volunteers when compared to the known transients in our field.

\begin{figure}
    \includegraphics[width=\columnwidth]{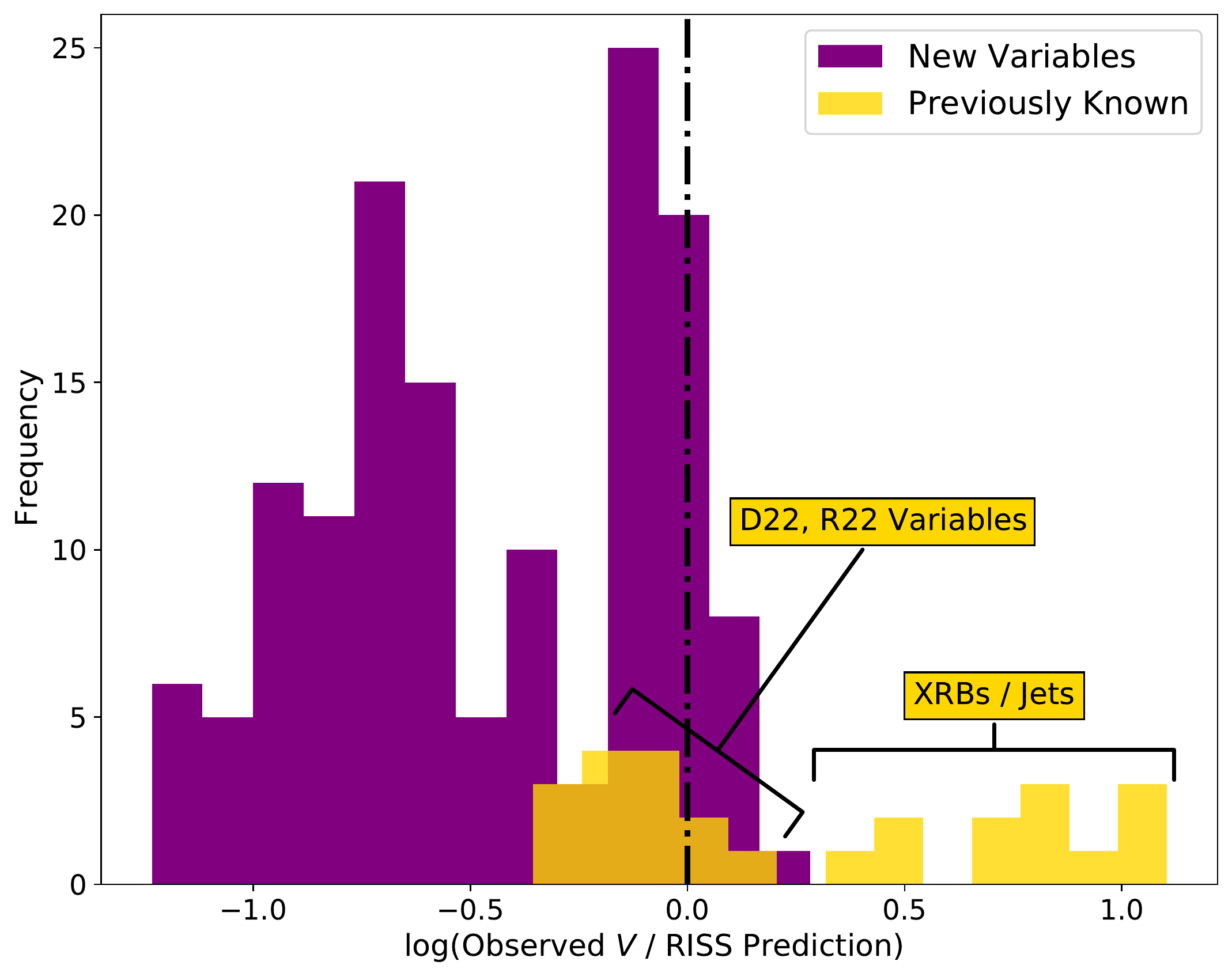}
    \caption{The ratio between observed $V$ and the variability predicted by \citet{Hancock2019}'s model for Refractive Interstellar Scintillation (RISS), for our sample of transients and variables. Most variability can be explained by this model of a scintillating, extragalactic source, apart from the known XRBs and jetted systems.}
    \label{fig:RISS}
\end{figure}

The majority of these 142 variables that our volunteers discover show long term variability, with light curves showing variations over weeks to months.
One cause of variability for extragalactic sources is scintillation through the Galactic interstellar medium (ISM). All of our observations are within latitude $|b| \lesssim 10$\degr  of the Galactic plane (see Table \ref{table:obs}) so this may be a large contributor to the variability seen in our sample. We can use the model from \cite{Hancock2019} to characterise the effect of refractive interstellar scintillation (RISS) for our set of variables, using an underlying H$\alpha$ map from \cite{2003ApJS..146..407F} to quantify the electron scattering along a given line of sight through the Galaxy. This model predicts the level of variability for a given radio frequency. We can directly compare this predicted maximum RISS-induced variability to our measured $V$ values, as seen in Figure \ref{fig:RISS}. For the majority (131) of our new variables we see that the predicted modulation is equal to or greater than that measured by \textsc{TraP} i.e. the observed variability is consistent with (though not exclusively explained by) the scintillation of light from AGN. By contrast, the known Galactic XRBs and their jets - whose variability is caused by shocks and particle acceleration - show variability that is much greater than what would be expected due to refractive scintillation. The predicted RISS variability is in some cases much higher than our observed $V$ values. This is likely caused by the heteregeneous sampling of our fields, the coarseness of the H$\alpha$ model grid and/or the assumptions of the model. Finally, we can calculate a weighted average time-scale of variation for our sources, which we find to be $t_0 = 8 \pm 4$ months, where the weights used are propagated through from the uncertainties in the underlying map and the quoted uncertainty is the standard deviation. This range of time-scales of variation at 1.28 GHz matches well with the length of typical observations for our XRB fields. Both these matching amplitudes and time-scales of variability provide evidence for the majority of our transients being scintillating AGN or other point-like extragalactic radio sources.

\subsection{Comparison to Target Sources}
\label{sec:targets}
\begin{figure*}
	\includegraphics[width=0.66\columnwidth]{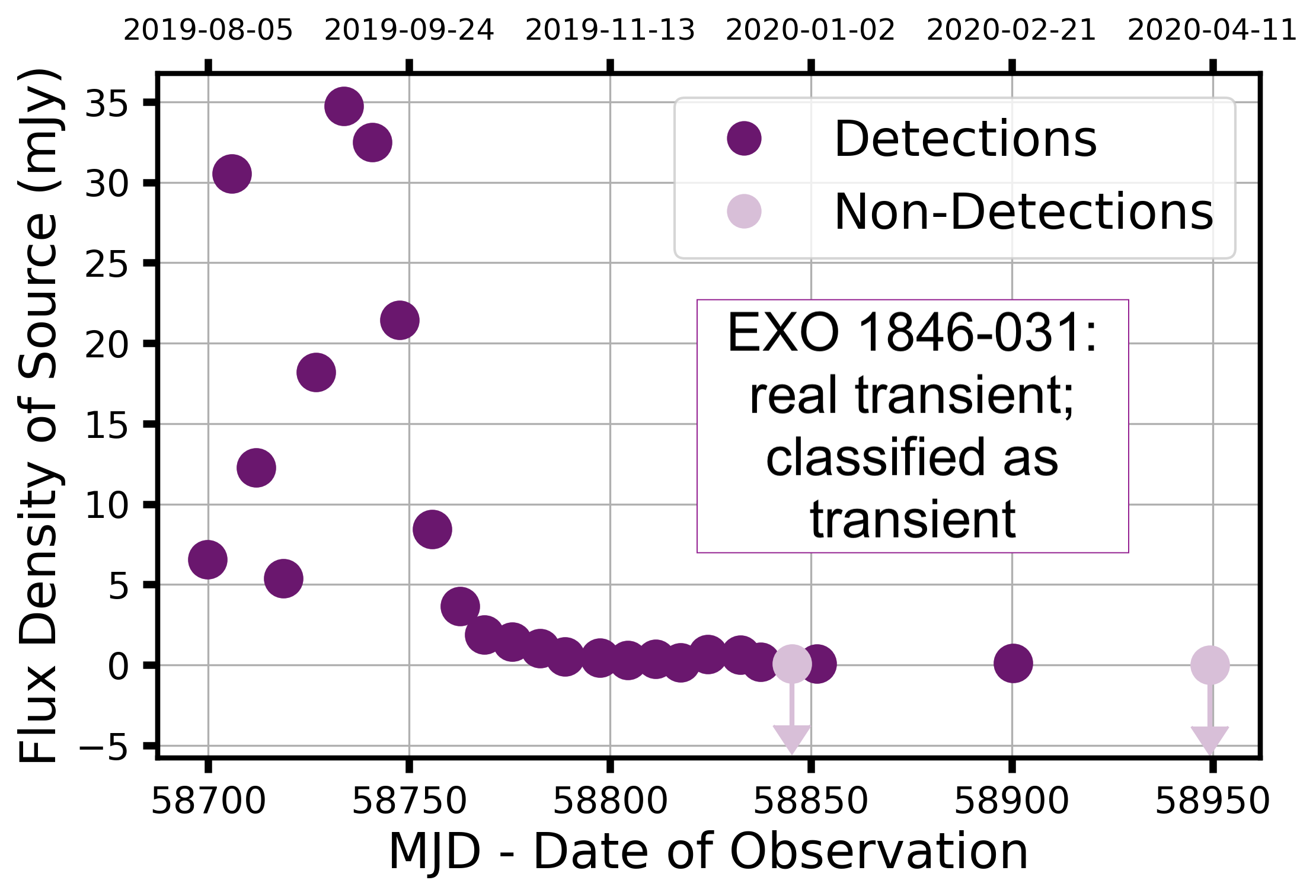}
	\includegraphics[width=0.66\columnwidth]{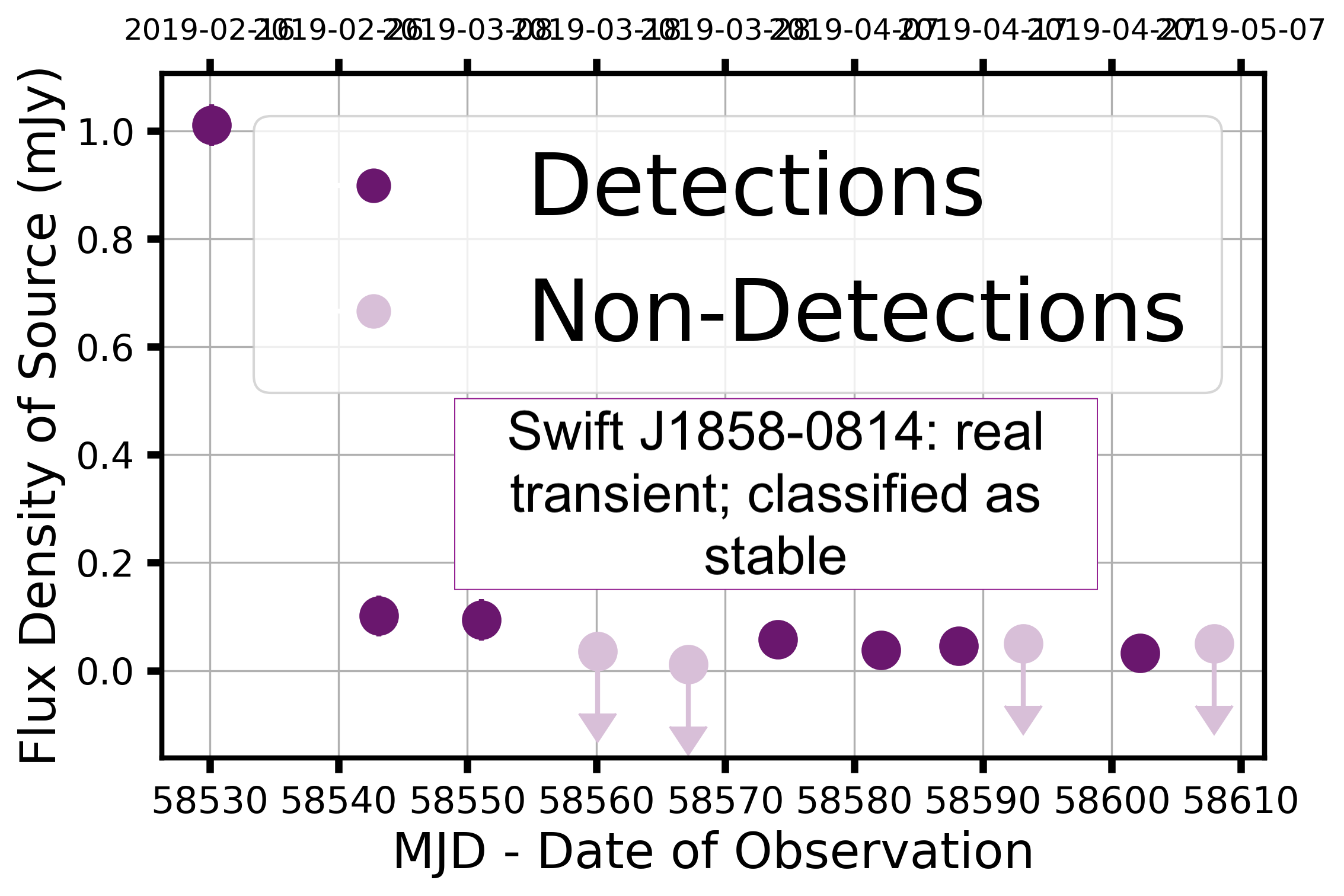}
	\includegraphics[width=0.66\columnwidth]{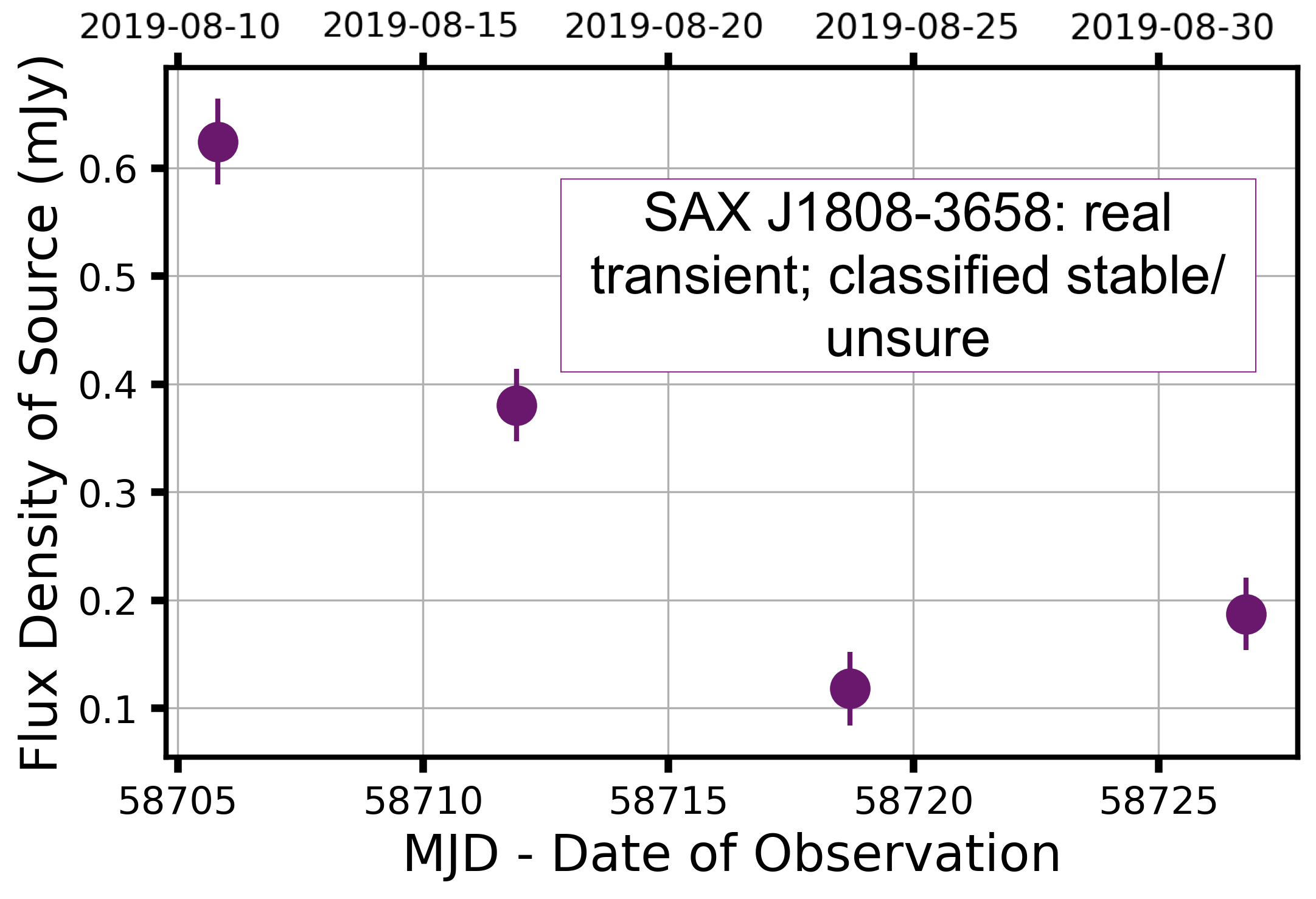}
    \caption{The \textsc{TraP} light curves of EXO 1846--031, \textit{Swift} J1858.6--0814 and SAX J1808.4--3658 (upper left, upper right and lower respectively). The foremost was classified as transient by citizen scientists whilst the others were not. We believe these to be caused by bad figure generation and lack of data points respectively.}
    \label{fig:XRBLCs}
\end{figure*}

Of all the 8874 sources classified in this project, there are 45 known variable/transient objects published in the literature, including the 11 XRBs listed in Table \ref{table:obs}.
Of the 11 XRBs, 9 are classified as transient by citizen scientists. The only two that are missed are \textit{Swift} J1858 and SAX J1808, whose light curves can be seen along with that of EXO 1846 for contrast in Figure \ref{fig:XRBLCs} - exactly as citizen scientists would have seen them. We can explain why \textit{Swift} J1858 was not classified as transient due to a combination of there being only one significantly bright data point, as well as the figure generation creating an overly large legend. Similarly, SAX J1808's misclassification can be understood as stemming from uncertainty surrounding so few data points (especially compared to other datasets). Indeed, SAX J1808 received 1 `Unsure' vote which, had it instead been for `Transient/Variable' would have pushed this subject above our classification threshold.

In addition to these central 11 sources, several of the XRBs also display discrete, resolved jet components (\textit{MAXI} J1820+070, \textit{MAXI} J1348--630, \textit{MAXI} J1848--015 and 4U1543--47; see Table \ref{table:obs}) that are counted as unique sources by the \textsc{TraP} as they become resolved and move away from the core of the jet. In total there are 6 moving, transient jet components that are detected by the \textsc{TraP} as unique, of which 3 are classified as variable by citizen scientists. Given that neither the software pipeline nor project classifications were designed to pick up moving point sources, it is interesting to note that there is room for unexpected discoveries even in such simple workflows.

\subsection{Previous commensal studies}
\label{sec:commensal}

One of the largest works on variable sources in MeerKAT images to date is that of \citetalias{Driessen2022}. In that work, 21 new long term variables (LTVs) are detailed, along with GX339--4,  the first MeerKAT transient MKT J170456.2--482100 and the known mode-changing pulsar PSR J1703--4851 (both described in \citealp{Driessen2020}). Of these 24, excluding the XRB discussed earlier,  MKT J170456.2−-482100, PSR J1703--4851 and 10 of the LTVs are missed by classifications i.e. only 11 are marked as transient/variable by volunteers. 
One reason for this is that in \citetalias{Driessen2022}, each light curve is binned by a factor of 10 (i.e. every 10 datapoints are represented by their uncertainty-weighted mean) to make long term variability more apparent, something not done here. Furthermore, here we made no use of a deep, stacked observation in order to follow sources through every epoch (see section \ref{sec:TraPdetails}'s details on the \texttt{expiration} parameter), something that will result in different light curve shape and in one case resulted in a source not being detected at all in our set of images. In general, the LTVs not identified by volunteers were generally fainter and have less smooth light curve evolution. This may give some insight into a bias of our method; it is easier to find light curves with long-term, clear evolution as opposed to more stochastic variability.

Similarly, when comparing to the work of \citetalias{Rowlinson2022} we see that the three variables found therein - NVSS sources J181849+062843, J181752+064638 and J182029+063419 - were recovered by citizen scientists. However, the three sources found in the `transient hunt' (where therein transient is defined as sources not detected in a deep observation of the field, see their section 3.1) were not identified by volunteers. As above we note that the variables recovered have smooth light curve evolution over long time-scales, whilst those not identified as transient are much fainter, with larger uncertainties and figures that are more `cluttered'.

Finally, the M dwarf SCR 1746--3214 is a radio transient that exhibits flares, serendipitously seen in early ThunderKAT data \citep{Andersson2022}. In order to assess what light curves citizen scientists were most comfortable with classifying, we provided two light curves of this source - one with more datapoints and an additional detection and one with only two detections and two upper limits (see Figure 3 in the above article). Interestingly, when provided with the shorter light curve of the source , the transient source was mostly classified as  `Unsure' or `Stable'. However, when given the full 
light curve volunteers correctly identify the subject as transient. We can perhaps use this to infer that citizen scientists were least unsure when classifying sources with more data points and less reliant on upper limits, as is the case for many of the new variables found in this study.

\section{Counterparts and Associations}
\label{sec:mw}

We make use of the MeerLICHT optical telescope to gain physical insight into possible source classes of our radio variables and transients. We use MeerLICHT for this due to its position in South Africa and its mission to follow the radio observations of MeerKAT, resulting in highly complementary spatial and temporal coverage of our radio sources (if observed at night). A typical observing schedule consists of 1 minute exposures of a given field, alternating between q-band (440--720nm) and each of 5 Sloan bands \textit{u}, \textit{g}, \textit{r}, \textit{i} and \textit{z}.  Whilst MeerLICHT operates in these 6 bands, here we use only the q-band due to its highest sampling rate and its broad wavelength coverage. We crossmatch with the MeerLICHT database at a 2\arcsec search radius - large enough to partially account for MeerKAT astrometric uncertainties (see \citetalias{Driessen2022}) and proper motion (e.g. of nearby stars) but not so large as to include many false matches - using the the uncertainty-weighted mean position of each of our candidates as returned by the \texttt{TraP}. Running this crossmatch returns 25 counterparts in the MeerLICHT database and 143 3$\sigma$ upper limits. This low rate of optical counterparts is not surprising as all of our XRB fields are within $~$10 degrees of the Galactic plane and so many optical counterparts may be heavily extincted.

\begin{figure*}
	\includegraphics[width=\textwidth]{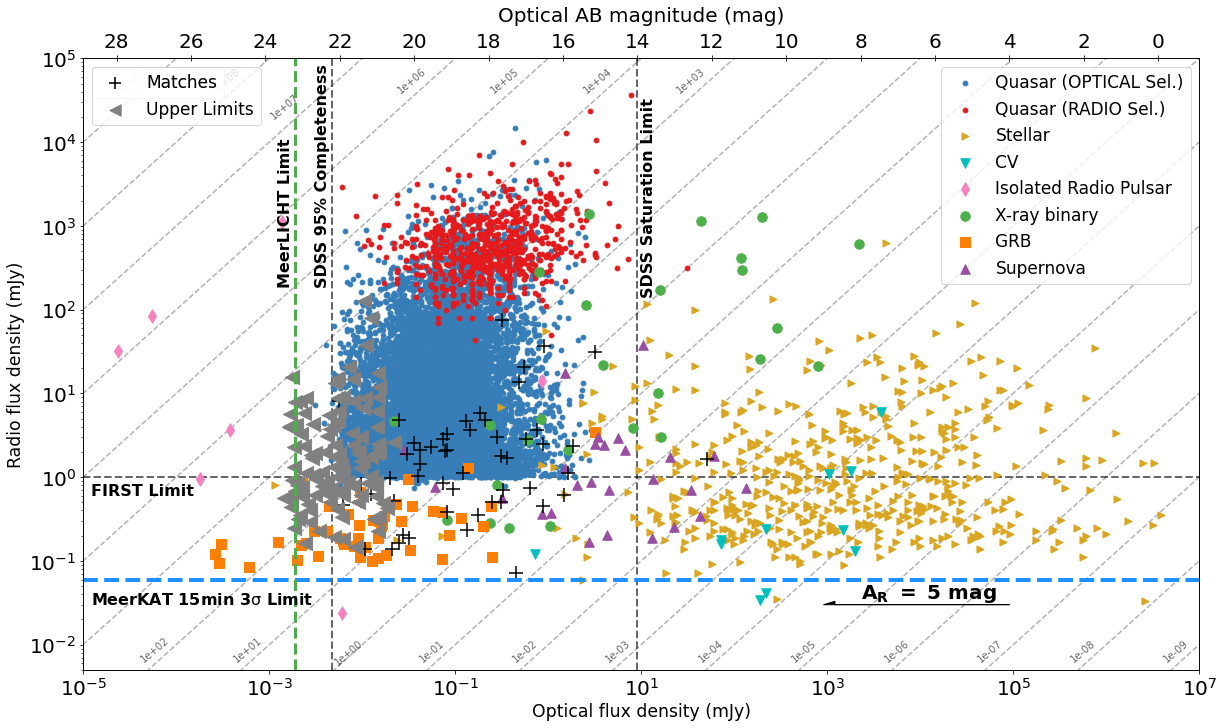}
    \caption{The mean optical and radio flux densities of our sample of radio variables, atop an underlying distribution of astrophysical classes \citep{Stewart2018}. Black crosses denote counterparts within the MeerLICHT database whilst grey triangles are upper limits. Diagonal lines denote a constant ratio between radio and optical flux density, whilst the $A_R$ marker indicates the horizontal displacement caused by 5 magnitudes of optical extinction. The majority of our radio sources are likely extragalactic as they overlap in parameter space with quasars and GRBs.}
    \label{fig:optical}

\end{figure*}

The radio-optical plane for our variables can be seen in Figure \ref{fig:optical}, with comparison source types from \cite{Stewart2018}\footnote{Code for reproducing this plot is available at \url{https://github.com/4pisky/radio-optical-transients-plot}}. The first thing to note is that the majority of these cross-matches exist in the region where extragalactic sources have been detected - either quasars or GRBs. If most of our candidates are extragalactic in nature this agrees with previous studies, who find the vast majority of radio variables are extragalactic \citep{Thyagarajan2011, 2021ApJ...923...31S}. There are a few sources overlapping the `stellar' region of the parameter space - these include the transients already reported by ThunderKAT (in \citealp{Driessen2020} and \citealp{Andersson2022}). It is important to note that by comparing to archival data in this way does not leave room for unknown astrophysical classes, however this still gives an indication of the overall distribution of multiwavelength counterparts.

\subsection{Highlights}

We also search for counterparts at other wavelengths and with pre-existing classifications, with the aid of pre-existing code available in \cite{laura_driessen_2021_4460968}. Crossmatching makes use of the \texttt{astroquery} package to search the SIMBAD\footnote{The Set of Identifications, Measurements and Bibliography for Astronomical Data, available online at \url{http://simbad.cds.unistra.fr/simbad/}} database \citep{Wenger2000} and several catalogues within Vizier\footnote{\url{https://vizier.cds.unistra.fr/viz-bin/VizieR}} \citep{2000A&AS..143...23O}. We again search at a 2\arcsec  radius to our radio sources. We searched for X-ray and gamma-ray counterparts to our radio variables. To do this we crossmatched with catalogs from the \textit{Fermi} \citep{Schinzel2014}, \textit{Chandra}, \citep{Evans2010}, \textit{Swift} \citep{Evans2020} and \textit{XMM-Newton} \citep{Traulsen2020} facilities. There were no counterparts for any of our sample, aside from the known transients of interest (i.e. the XRBs).
Below we make note of a few interesting objects returned in our search with IR or radio detections, or otherwise known counterparts.

\subsubsection{OH Maser - BfS 80}

\begin{figure}
	\includegraphics[width=\columnwidth]{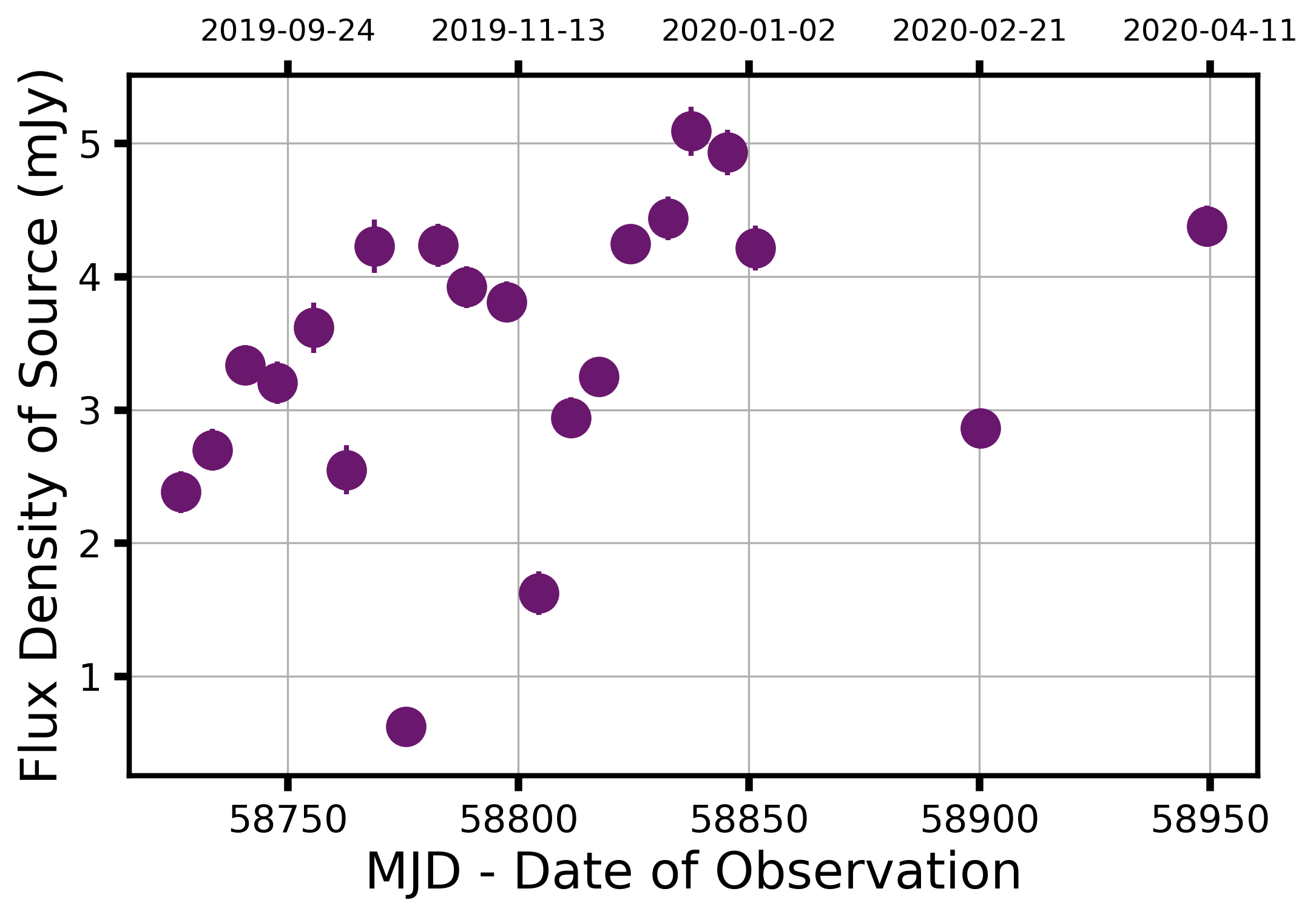}
 	\includegraphics[width=\columnwidth]{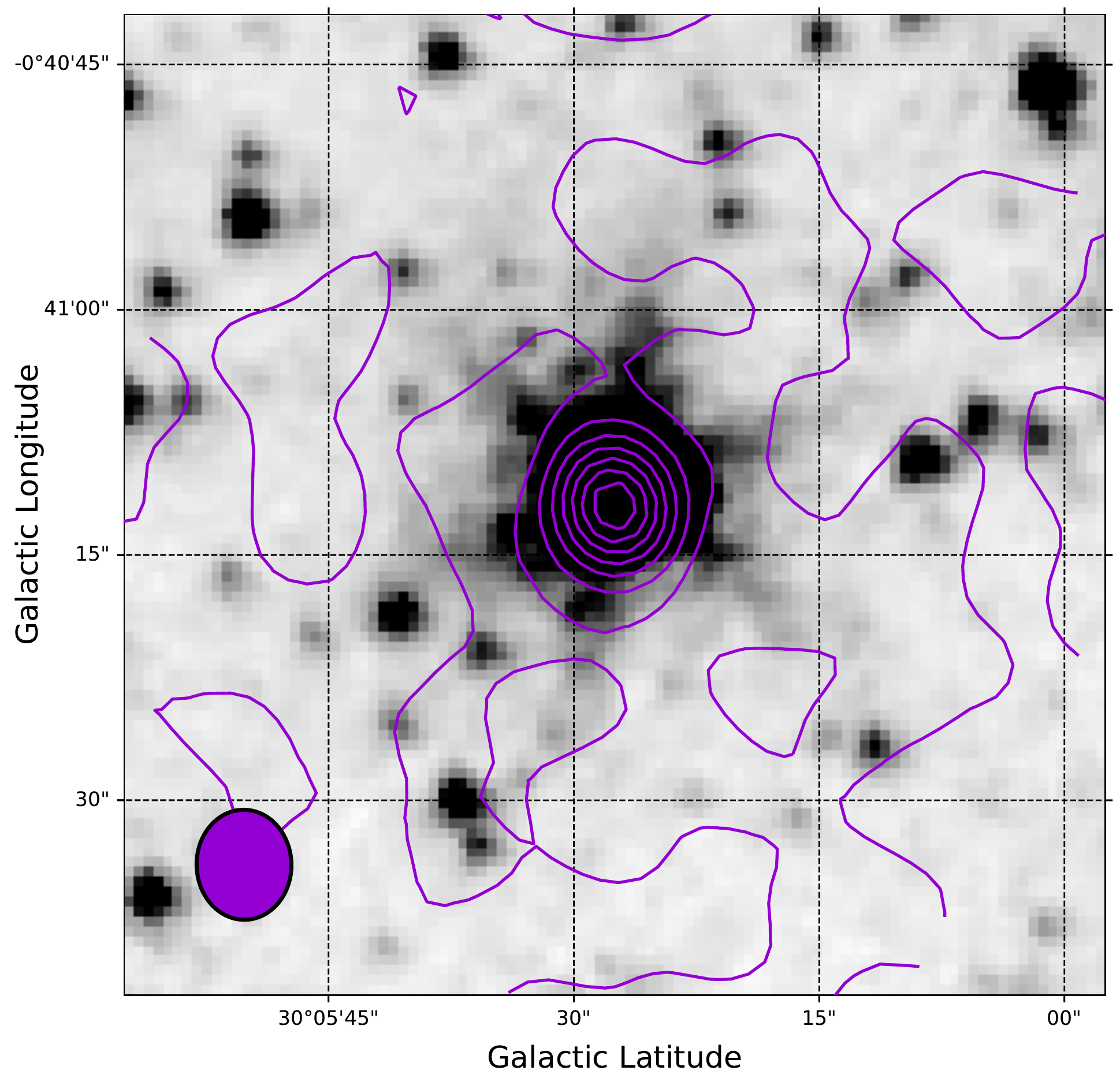}
    \caption{\textbf{Upper:} Light curve of OH Maser 30.1--0.7 picked up by volunteers. There were 4 epochs in which this source was not detected prior to this observed variability. \textbf{Lower:} Overlay of MeerKAT radio contours over Spitzer imaging from the GLIMPSE survey \citep{benjamin2003}. Contours are spaced linearly in 0.5 mJy increments from -0.5 to 3.5 mJy.} 
    \label{fig:OHMaser}
\end{figure}

We detected transient emission from a source in the EXO 1846 field which volunteers confidently classified as transient. Cross-matching reveals that the source in question is a known maser star OH 30.1--0.7 otherwise known as V1362 Aql, with radio observations stretching back almost 50 years \citep{Evans1976}. This asymptotic giant branch (AGB) star is heavily dust-obscured at optical wavelengths, but very bright at IR wavelengths - $W1 = 6.9 \pm 0.1$ mag, or 279 Jy at 25 $\mu$m \citep{Cutri2013, Gonidakis2014}. A comparison of a MeerKAT radio detection (contours) and Spitzer Glimpse imaging can be seen in Figure \ref{fig:OHMaser}, alongside its light curve. AGB stars are post-main sequence systems, whose low surface temperatures, radii of several hundred times the solar radius and stellar pulsations give rise to strong winds, expelling as much as $\sim 10^{-5} M_{\odot}$yr$^{-1}$ \citep{Hofner2018}. These winds create an oxygen-rich, dusty circumstellar environment that generate masers at 1612 and 1667 MHz as infrared photons pump OH molecules formed through photodissociation of water by interstellar radiation.

The cause of the variability of OH 30.1--0.7 is not clear. Perhaps the OH maser emission is varying, due to stellar pulsations. The derived stellar period from the General Catalog of Variable Stars is $\sim$1730d \citep{Samus2017}, with more recent estimates from the WISE W1 and W2 bands at $1950 \pm 50$ and $1520 \pm 20$ days respectively \citep{Groenewegen2022}. These time-scales are much longer than the variability seen in the radio light curve here (of order a few months), so it is not clear if stellar pulsations are responsible for variable maser emission (averaged across the entire L-band from Jy to mJy levels). There could also be inhomogeneities at the site of the maser emission. The second cause for variability could be due to binary interactions - the system likely has a companion, as inferred from ALMA CO data in \citet{Decin2019}. This, combined with the lack of any optical counterpart implies the system could be a dusty (D-type) symbiotic binary - all D-type symbiotics host Mira stars \citep{Whitelock1987}. These binaries consist of a windy red giant and a smaller companion onto which material is shed \citep{Allen1984}. Radio emission has been seen from symbiotic binaries \citep{Seaquist1984} and is known to vary in some sources, typically interpreted as optically thick emission from an inhomogeneous region in the red giant's wind that is ionised by its companion \citep{Seaquist1988}. The nature of OH 30.1--0.7's companion is as yet unknown due to the heavy extinction in this region and without evidence of high ionisation (e.g. HeII or [OIII]) we cannot claim that this is a symbiotic system.
There is overlap between maser systems and Mira-type symbiotic binaries \citep{Seaquist1995} and so the observed variability could be due to a combination of emission mechanisms discussed, or perhaps something else entirely. One thing to note is that there were four observations of this field prior to the initial data point of the light curve - i.e. there are 4 non-detections before the system reached $\sim$3mJy as seen. Future radio observations could help determine what the nature of the variability is (including the initial non-detections), perhaps combined with spectroscopic searches for the nature of the companion (e.g. if it is a white dwarf, or for evidence for ionisation).

\subsubsection{Pulsar - BfS 20}

\begin{figure}
	\includegraphics[width=\columnwidth]{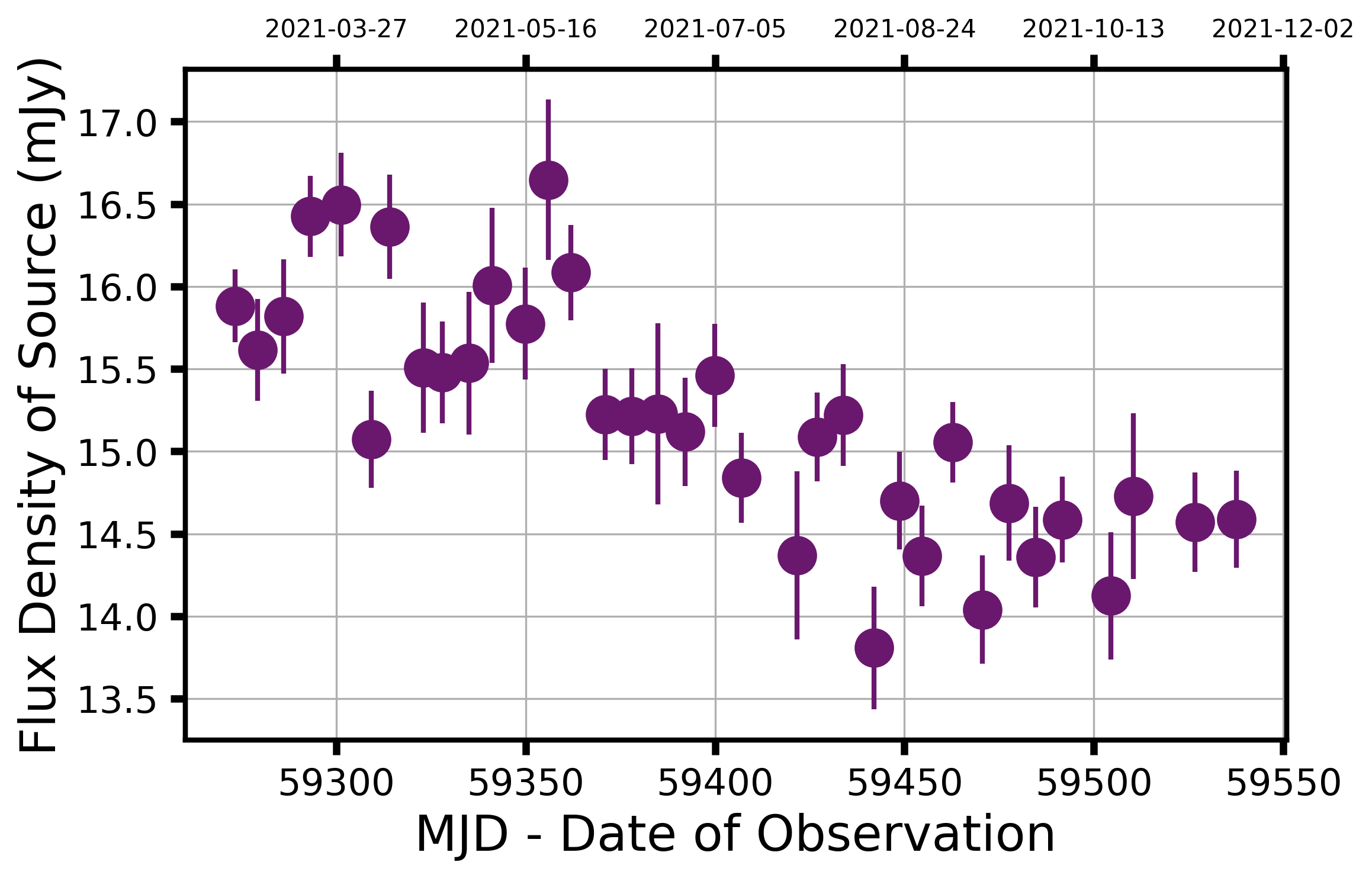}
    \caption{Light curve of PSR B1845--01 with MeerKAT. 7 of 10 volunteers voted for this as a transient/variable source. The low amplitude variability observed over a time-scale of 100s of days is consistent with RISS.} 
    \label{fig:PSR}
\end{figure}

Pulsar PSR B1845-01 (J1848--0123), whose light curve can be found in Figure \ref{fig:PSR}, was seen in the field surrounding XRB \textit{MAXI} J1848. In our MeerKAT observations we measure a mean flux density of $\sim$15 mJy, in close agreement with that measured by the Parkes radio telescope recently (15.2 $\pm$ 0.9 mJy; \citealp{Johnston2018}). However, reports of this pulsar's flux density have been as low as 8.9 $\pm$ 0.9 mJy at 1400 MHz \citep{Hobbs2004}, indicating long term changes in the received brightness from the source. The pulsar is brighter at lower frequencies (e.g. measured to be $79 \pm 6$ mJy at 408 MHz by \citealp{Lorimer1995}), with a spectral index $\alpha$ - defined at frequency $\nu$ as flux density $F \propto \nu^\alpha$  - of $-1.3 \pm 0.3$ \citep{2020ApJ...892...76M}.


The observed variability is consistent with RISS (see section \ref{sec:results}), with a predicted modulation greater than the measured $V \sim 0.02$ and the observed time-scale of variation matching the estimated $11 \pm 2$ months. Furthermore, this pulsar has a relatively high dispersion measure (DM) of $159.1 \pm 0.2$ pc cm$^{-3}$ (\citealp{2020ApJ...892...76M}; compared to typical DMs of order 10s of pc cm$^{-3}$ for pulsars with $|b| > 25\degr$; \citealp{Manchester2005}), which is known to be linked to long time-scale, low amplitude refractive scintillation \citep{Stinebring1990, Stinebring2000}. Similarly, the pulsar's location in the Galactic plane is in keeping with its radio emission traversing a large free electron content, hence the high DM and clear scintillation. We note that B1845-01's spin period of $\sim$0.65943s is much shorter than that of our observations (typically 15 minute epochs consisting of 8s correlator sampling) so this cannot be contributing to the observed variability.


This pulsar adds to the diverse range of behaviours seen in pulsars spotted by MeerKAT in imaging data. Similar examples include the mode changing pulsar observed in the GX339--4 field (see \citetalias{Driessen2022}) and, in the most extreme case, one of the slowest pulsars discovered \citep{Caleb2022}.

\subsubsection{VLASS1 J181955.28+074418.7 - BfS 146}

\begin{figure}
	\includegraphics[width=\columnwidth]{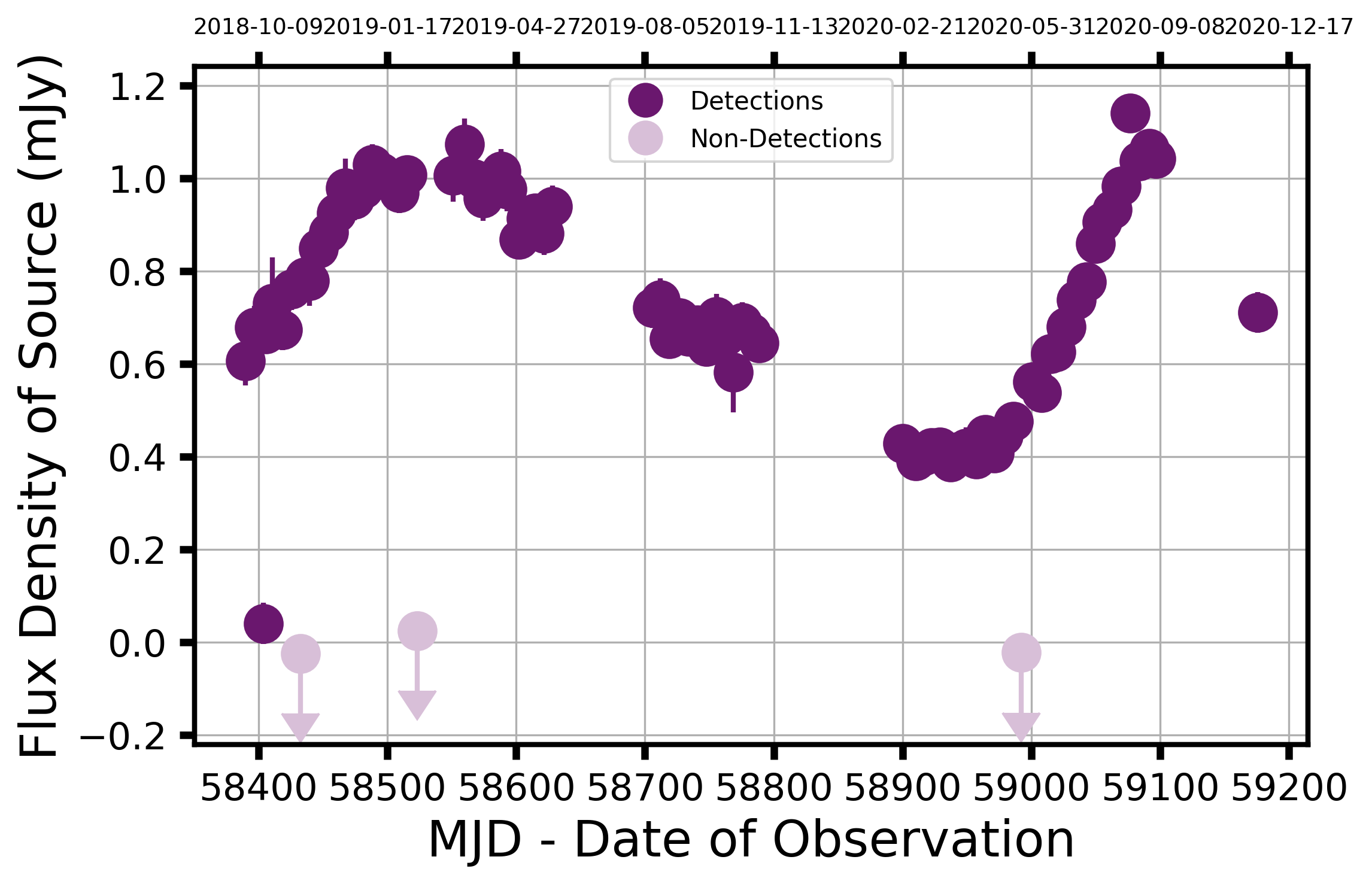}
    \caption{Light curve of source VLASS1 J181955.28+074418.7, showing smooth, near sinusoidal variations. The epochs denoted upper limits are in error and due to not filtering out low quality images in our pre-processing.} 
    \label{fig:AGN}
\end{figure}

Radio source VLASS1 J181955.28+074418.7 is an object spotted by our volunteers not only in our aggregated results (scoring 7/10 votes as a transient/variable) but also in our Talk board\footnote{\url{https://www.zooniverse.org/projects/alex-andersson/bursts-from-space-meerkat/talk/4567/2263526}}. This source is in the field of MAXI J1820 but outside the 0.5 degree radius set by \citetalias{Rowlinson2022}. Its light curve can be seen to vary very smoothly in Figure \ref{fig:AGN}, where it is worth noting that the non-detections are all due to drops in data quality in those images, which should be filtered out in future work. This variation is not similar to any nearby source in our data, nor is it correlated with PSF size or shape. There is a counterpart to this source in the Very Large Array's Sky Survey \citep[VLASS QuickLook Epoch 1;][]{Lacy2020, 2021ApJS..255...30G}, with a 3 GHz flux density of $1.9 \pm 0.3$mJy.  There are no counterparts to this source in any higher energy band, despite this field being our furthest from the Galactic plane ($b \sim 10\degr$). This kind of source is very typical of our sample - scant extra data but an intriguing radio light curve. Further information at other wavelengths would help determine source type, as would simultaneous radio data - e.g. to determine a precise spectral index $\alpha$ and see if this points towards the source being an AGN or a pulsar. Our RISS analysis (see section \ref{sec:results}) produces a ratio of 1.17 between the observed $V$ and predicted scintillation amplitude. This could indicate the source has some intrinsic variation however the scintillation parameters were not exhaustively tested, so this ratio could be explained by an incorrectly assumed distance to, or relative motion between, the model scattering screen and the observer.

\section{Discussion}
\label{sec:disc}

Using citizen science to discover transients is a fruitful endeavour, as shown by both the uptake of our Zooniverse project and the novel results produced. For our volunteers, the project provides new experiences with a branch of observations perhaps less familiar to their perception of astronomy (e.g. compared to galaxy morphologies or solar physics). The uptake of our project demonstrates a clear appetite for further development of citizen science for radio transients, as evidenced by >1000 citizen scientists applying their time to our science case. The time taken to individually check several years worth of data is much greater than the 3 months taken by our volunteers, so citizen scientists help project scientists analyse their observations more efficiently. In this study we have been able to recover or discover a broad range of astrophysical transients that occur over several orders of magnitude in time-scale, so the science results discussed here also give merit to developing BfS:MKT further. These results include flaring from stellar systems, the discrete and compact jets of XRBs, maser emission, pulsars, the long-term variation in likely AGN and potentially new source classes. Given these successes, we plan to launch a second wave of classifications to the Zooniverse site shortly, using different data from surveys on MeerKAT to explore the range of parameter space for both citizen and project scientists. Of course this method is not restricted to MeerKAT data and any such set of (radio) images and light curves could be analysed in this way.

However, citizen science is not without its challenges and we can understand some of these by comparing our findings to those of previous work from the ThunderKAT team - namely \citetalias{Driessen2022} and \citetalias{Rowlinson2022}, as well as the published XRB work (see Table \ref{table:obs}). The first thing to note is that not all of the transients from \citetalias{Driessen2022} and \citetalias{Rowlinson2022} were recovered here. Some of this can be explained by the difference in pre-processing of images and \textsc{TraP} light curves (discussed in \ref{sec:commensal}). However, several of the transients missed are due to how we have produced subject images for the Zooniverse project. For example, the two XRBs not classified as transients can be explained due to the lack of clarity in Figure \ref{fig:XRBLCs} (see section \ref{sec:targets}). Similarly, the image of source MKT J182015.5+071455 (\citetalias{Rowlinson2022}'s source 2) provided to volunteers was automatically scaled to the brightest pixel in the image, not be the central source of interest. As such, the source appears not to be present and was classified mostly as an artefact. We will use these issues to improve our procedure for future batches of data e.g. by manually setting the pixel scale in images, altering legend sizes etc.

Aside from these pre-processing issues, there are still some known transients not recovered, including the pulsars in \citetalias{Driessen2022} and \citetalias{Rowlinson2022}. When we compare the light curves of these pulsars to variables that are recovered we see a clear trend -  recovered transients show long term, smooth variations, and are typically brighter, with smaller uncertainties, resulting in clearer patterns on display to scientists. By contrast, the faint, transient pulsar light curves show very `noisy' light curves, with less eye-catching patterns, despite being precisely the kinds of transients we want to discover. \citetalias{Rowlinson2022}'s pulsar received 3 votes as a transient/variable and 1 unsure classification - this could be due to the heterogeneous nature of our classifiers, or it could reflect the uncertainty surrounding a less clear pattern in a light curve. As mentioned in section \ref{sec:commensal}, we provided two light curves of \cite{Andersson2022}'s M dwarf to volunteers, one with force fit measurements (more data points and an additional detection) and one without. The light curve with more data points passed our threshold of 0.4, whilst the latter did not, perhaps indicating that classifiers are more comfortable with longer light curves. These unrecalled transients give insight into the limitations of this dataset: our sample of transients and variables is likely biased towards brighter, slower evolving objects that occur in our most sampled fields. We hope to alleviate some of this bias in future Zooniverse runs by emphasising the use of non-detections and by encouraging volunteers to label things as transient. We could also implement the transients and variables discussed in this work into the Field Guide for volunteers as more examples of the types of sources for which we are looking.

 We can quantify how scalable our method is by comparing our observations to the MeerKAT MIGHTEE survey. MIGHTEE's observations produce $\sim6000$ sources per square degree on the sky \citep{2022MNRAS.509.2150H}, for which our \textsc{TraP} processing would produce a light curve.  For comparison, in this work we produce $\sim1000$ subjects from the ThunderKAT fields that are devoid of large diffuse structures or exceedingly bright sources. If we assume that only the same 1000 volunteers contribute to all future data releases as with this study, at the same classification rate, then it would take approximately 60 days to receive 10 classifications on every source at MIGHTEE's sensitivity ($1\sigma$ RMS noise $\sim$ 1 $\mu$Jy), per square degree. Multiplying this across the sky coverage of just the ThunderKAT fields used in this study ($\sim1.5$ square degrees over 11 fields) results in a required volunteer classification time of 990 days. This is far greater than the time taken for the observations (e.g. of order tens of 8-hour epochs) and which is needed to image the data and process them to form light curves. If the volunteer results were only analysed when all classifications were finished then this would also be too late for real-time follow-up of transients. To bring the classification time to that of this work (90 days) would require an order of magnitude more volunteers, which is achievable for Zooniverse projects, particularly when disseminated widely. For example, \textit{Galaxy Zoo} variants receive many 10s of thousands of volunteers, whilst \textit{Gravity Spy} has had over 30,000 participants. However, for any considerable survey area the overall time required would again balloon to far larger time-scales than reasonable, particularly if we process datasets and classifications in batches once observations are complete. Finally, observations at the 8s integration time for MeerKAT produce far fewer sources to classify compared to deeper images, however the imaging and processing time prior to volunteer classification increases hugely so the overall time-scale remains long. So this kind of analysis will not scale easily to the most sensitive observations of MeerKAT fields (at 1.28 GHz), let alone those expected from the SKA, ngVLA \citep{Hallinan2019} or DSA-2000 \citep{Selina2018}.


One way to alleviate this data deluge might be to develop machine learning methods to remove `bogus' and `boring' sources in favour of the rarer variables and transients for which we are searching, as has been done for e.g. supernovae and galaxy morphologies \citep{Wright2017, Walmsley2022}. Active learning, where humans feed back to machine learning techniques in order to prioritise sources of interest and optimise precious human attention, has been shown to uncover unique light curves \citep{Ishida2021} and radio morphologies \citep{Lochner2023}, and are able to optimise the volunteer classification of optical galaxies \citep{Walmsley2022a}. We are currently applying \citet{Lochner2021}'s \textsc{Astronomaly} active learning framework to the data presented in this work in the hopes that we can use the combined power of human classifiers and machine processing to extract the most science from our wealth of data (Andersson et al. in prep).  


\section{Conclusions}
\label{sec:conclude}

In this work we have presented the first citizen science project for finding transients in image plane radio surveys. The uptake of the project was very strong, with $>1000$ volunteers taking part, demonstrating a healthy appetite for further Zooniverse data releases. We were also able to use the known transients in our fields to understand some reasons why interesting sources may be missed and will fold this learning through to future iterations of the project. Citizen and project scientists uncovered a large sample of interesting transient and variable sources, some of which we may have not uncovered were it not for our volunteers' dedication. We provide the full catalogue of 168 radio transients and variables, the largest catalogue of candidate radio variables to date. This includes links to images and light curves, and we encourage others to follow up these sources, and additional future catalogues that this project will deliver. We made use of archival multiwavelength data, including the MeerLICHT telescope, to help classify the systems found. The sources found span a broad range of physical phenomena including pulsars, radio-loud stars, XRBs and a large set of AGN - likely varying due to scintillation. These results demonstrate the wealth of science possible with new radio facilities.  Finally, we hope to use volunteer classifications to develop anomaly detection algorithms, with an eye towards current and future surveys such as the SKA.

\section*{Acknowledgements}

The authors would like to thank our reviewer Matthew Bailes for his helpful comments that have improved this study. We also thank Patricia Whitelock for her helpful discussion points around OH masers, Mira variables and symbiotic variables. 
AA acknowledges the support given by the Science and Technology Facilities Council through a STFC studentship. 
CJL acknowledges support from the Alfred P. Sloan foundation.
FC acknowledges support from the Royal Society through the Newton International Fellowship programme (NIF/R1/211296).
LND acknowledges support from the European Research Council (ERC) under the European Union's Horizon 2020 research and innovation programme (grant agreement No 694745).
PJG acknowledges support from NRF SARChI Grant 111692.
MeerKAT is operated by the South African Radio Astronomy Observatory (SARAO), which is a facility of the National Research Foundation, an agency of the Department of Science and Innovation. 

This publication uses data generated via the \url{Zooniverse.org} platform, development of which is funded by generous support, including a Global Impact Award from Google, and by a grant from the Alfred P. Sloan Foundation.

We thank the SARAO staff involved in obtaining the MeerKAT observations.
We acknowledge the use of the ilifu cloud computing facility – \url{www.ilifu.ac.za}, a partnership between the University of Cape Town, the University of the Western Cape, the University of Stellenbosch, Sol Plaatje University, the Cape Peninsula University of Technology and the South African Radio Astronomy Observatory. The ilifu facility is supported by contributions from the Inter-University Institute for Data Intensive Astronomy (IDIA – a partnership between the University of Cape Town, the University of Pretoria, the University of the Western Cape and the South African Radio astronomy Observatory), the Computational Biology division at UCT and the Data Intensive Research Initiative of South Africa (DIRISA).
The MeerLICHT consortium is a partnership between Radboud University, the University of Cape Town, the Netherlands Organisation for Scientific Research (NWO), the South African Astronomical Observatory (SAAO), the University of Oxford, the University of Manchester and the University of Amsterdam, in association with and, partly supported by, the South African Radio Astronomy Observatory (SARAO), the European Research Council and the Netherlands Research School for Astronomy (NOVA).

This research has made use of the SIMBAD database, operated at CDS, Strasbourg, France \citep{Wenger2000}, as well as the ATNF Pulsar catalog \citep{Manchester2005} available at \url{https://www.atnf.csiro.au/research/pulsar/psrcat/}.

This research has made use of Astropy, an Astronomy-based, community-developed Python package \citep{astropy:2013, astropy:2018, astropy:2022}.

\section*{Data Availability}

ThunderKAT raw data are available on the SARAO archive (\url{https://archive.sarao.ac.za/}).
The data for each light curve mentioned in Table \ref{tab:results} and throughout the text will be available at \url{https://github.com/AnderssonAstro/BfS-MKT-Analysis} and we encourage readers interested in particular sources to investigate them in further detail, giving credit to this work.
The data used to produce our figures will also be in said repository, excluding personal data from volunteers, 
which can be shared on reasonable request to the authors.



\bibliographystyle{mnras}
\bibliography{Refs.bib} 



\appendix

\section{Table of Results}

\onecolumn
\label{tab:results}

\begin{longtable}{|c|c|c|c|c|c|c|c|c|c|c|}

\caption{Table of transients and variables found by volunteers during BfS:MKT, with their associated positions, median flux densities ($F_{\mathrm{med}}$) and variability parameters $\eta$ and $V$. The observed date given here is that in which the source was detected at highest S/N by the \textsc{TraP}, whilst the distance recorded is how far an entry is from the pointing centre of that observation. The Transient Fraction (TF) is the fraction of classifications as a transiet/variable, from 10 volunteers. Note that their Zooniverse subject ID is a unique identifier from which one can find the image and light curve online and freely accessible at \url{https://www.zooniverse.org/projects/alex-andersson/bursts-from-space-meerkat/talk/subjects/ID}, replacing \texttt{ID} with the numeric value below. This table will be available in machine-readable form online.}\\

\hline
Name & Subject ID & Right Ascension & Declination & Date Obs. & TF & $F_{\mathrm{med}}$ & $\eta$& $V$ & Distance & Known? \\
& & (\degr) & (\degr) & & & (mJy) & & & (\arcmin) & \\
\hline
\endfirsthead

\multicolumn{11}{l}%
{{\bfseries \tablename\ \thetable{} -- continued from previous page}} \\
\hline
Name & Subject ID & Right Ascension & Declination & Date Obs. & TF & $F_{\mathrm{med}}$ & $\eta$& $V$ & Distance & Known? \\
& & (\degr) & (\degr) & & & (mJy) & & & (\arcmin) & \\
\hline
\endhead

\hline \multicolumn{11}{|r|}{{\textit{Continued on next page}}} \\ \hline

\endfoot

\hline \hline
\endlastfoot

    \hline

        BfS 0 & 70785323 & 255.424 & -48.776 & 14/04/2018 & 0.4 & 13.2 & 83.7 & 0.05 & 11.2 & 0 \\  
        BfS 1 & 70785357 & 255.364 & -48.970 & 14/04/2018 & 0.9 & 11.2 & 395.4 & 0.11 & 17.3 & 1 \\  
        BfS 2 & 70785384 & 254.710 & -48.877 & 14/04/2018 & 0.5 & 4.7 & 26.8 & 0.07 & 39.7 & 0 \\  
        BfS 3 & 70785398 & 255.441 & -48.675 & 14/04/2018 & 0.6 & 0.7 & 1.8 & 0.11 & 12.5 & 1 \\  
        BfS 4 & 70785401 & 255.369 & -48.499 & 14/04/2018 & 0.4 & 0.4 & 0.6 & 0.14 & 22 & 1 \\  
        BfS 5 & 70785440 & 255.156 & -48.946 & 14/04/2018 & 0.5 & 1.7 & 6.6 & 0.09 & 23.6 & 1 \\  
        BfS 6 & 70785461 & 255.634 & -48.850 & 14/04/2018 & 0.9 & 0.6 & 4.3 & 0.19 & 4.6 & 0 \\  
        BfS 7 & 70785641 & 255.291 & -48.597 & 14/04/2018 & 0.7 & 3.9 & 100.6 & 0.15 & 20.1 & 1 \\  
        BfS 8 & 70785723 & 255.117 & -48.429 & 14/04/2018 & 0.6 & 0.6 & 1.1 & 0.11 & 31.9 & 1 \\  
        BfS 9 & 70785821 & 256.988 & -48.972 & 14/04/2018 & 0.4 & 8.8 & 130.5 & 0.09 & 51.8 & 0 \\  
        BfS 10 & 70785881 & 255.917 & -48.670 & 14/04/2018 & 0.4 & 0.5 & 0.9 & 0.11 & 11.1 & 1 \\  
        BfS 11 & 70785898 & 255.706 & -48.790 & 01/02/2020 & 0.7 & 2.5 & 22427.6 & 1.86 & 0 & 1 \\  
        BfS 12 & 70785951 & 255.983 & -48.932 & 14/04/2018 & 0.6 & 3.6 & 62.6 & 0.13 & 13.9 & 1 \\  
        BfS 13 & 70785952 & 255.557 & -48.560 & 14/04/2018 & 0.8 & 1.4 & 3.6 & 0.08 & 15 & 1 \\  
        BfS 14 & 70785973 & 256.443 & -48.806 & 21/03/2020 & 0.5 & 0.2 & 0.6 & 0.23 & 29.2 & 1 \\  
        BfS 15 & 70786097 & 289.084 & 10.836 & 28/02/2021 & 0.9 & 0.8 & 7.6 & 0.32 & 18 & 0 \\  
        BfS 16 & 70786101 & 288.964 & 10.940 & 05/04/2021 & 0.4 & 0.2 & 8.7 & 0.49 & 9.8 & 0 \\  
        BfS 17 & 70786128 & 288.798 & 10.946 & 24/07/2020 & 0.6 & 37 & 66097.3 & 1.51 & 0.3 & 1 \\  
        BfS 18 & 70786366 & 288.400 & 10.354 & 08/12/2018 & 0.4 & 8.1 & 8002.1 & 0.24 & 42.3 & 0 \\  
        BfS 19 & 70786703 & 282.208 & -1.497 & 08/11/2021 & 0.4 & 4.8 & 37 & 0.46 & 0.3 & 1 \\  
        BfS 20 & 70786771 & 282.098 & -1.400 & 28/02/2021 & 0.7 & 15.2 & 5 & 0.05 & 8.6 & 0 \\  
        BfS 21 & 70786879 & 282.207 & -1.499 & 08/11/2021 & 0.9 & 15.7 & 984.2 & 0.67 & 0.4 & 1 \\  
        BfS 22 & 70786931 & 282.080 & -1.679 & 04/10/2021 & 0.5 & 80.4 & 12.5 & 0.02 & 13.6 & 0 \\  
        BfS 23 & 70789583 & 271.374 & -30.120 & 22/05/2021 & 0.4 & 9.1 & 23 & 0.02 & 36.4 & 0 \\  
        BfS 24 & 70789643 & 270.424 & -29.801 & 31/07/2021 & 0.6 & 0.5 & 1 & 0.07 & 17.6 & 0 \\  
        BfS 25 & 70789787 & 271.225 & -29.740 & 07/05/2021 & 0.6 & 1.9 & 2.9 & 0.03 & 24.7 & 0 \\  
        BfS 26 & 70789902 & 270.687 & -30.534 & 12/06/2021 & 0.6 & 2.1 & 2.1 & 0.02 & 42.6 & 0 \\  
        BfS 27 & 70789937 & 271.029 & -30.293 & 15/08/2021 & 0.5 & 127.3 & 99.8 & 0.02 & 31.2 & 0 \\  
        BfS 28 & 70790055 & 270.934 & -29.951 & 12/06/2021 & 0.5 & 1.6 & 2.4 & 0.04 & 11.7 & 0 \\  
        BfS 29 & 70790141 & 270.681 & -29.466 & 12/06/2021 & 0.4 & 2 & 3.8 & 0.03 & 22 & 0 \\  
        BfS 30 & 70790174 & 270.699 & -29.561 & 07/05/2021 & 0.8 & 2.4 & 3.5 & 0.03 & 16.2 & 0 \\  
        BfS 31 & 70790384 & 271.383 & -30.411 & 12/06/2021 & 0.4 & 6.2 & 20.9 & 0.03 & 47.7 & 0 \\  
        BfS 32 & 70790438 & 271.424 & -30.144 & 07/05/2021 & 0.5 & 8 & 11.9 & 0.02 & 39.3 & 0 \\  
        BfS 33 & 70790547 & 270.539 & -29.637 & 19/06/2021 & 0.7 & 1.9 & 8.5 & 0.05 & 16.2 & 0 \\  
        BfS 34 & 70790666 & 270.559 & -29.467 & 05/09/2021 & 0.4 & 3 & 1.8 & 0.02 & 24 & 0 \\  
        BfS 35 & 70790672 & 271.167 & -29.189 & 19/06/2021 & 0.5 & 6.2 & 36.7 & 0.04 & 43.7 & 0 \\  
        BfS 36 & 70790754 & 270.762 & -29.830 & 15/05/2021 & 0.4 & 0.1 & 13026.9 & 2.38 & 0.2 & 1 \\  
        BfS 37 & 70790813 & 270.399 & -29.923 & 12/07/2021 & 0.4 & 1 & 2.2 & 0.05 & 19.7 & 0 \\  
        BfS 38 & 70790848 & 270.833 & -29.289 & 23/10/2021 & 0.5 & 0.8 & 2.4 & 0.07 & 32.5 & 0 \\  
        BfS 39 & 70790883 & 271.545 & -30.141 & 27/05/2021 & 0.7 & 2.1 & 4.5 & 0.03 & 44.9 & 0 \\  
        BfS 40 & 70790918 & 270.650 & -29.852 & 05/06/2021 & 0.5 & 2.8 & 14.3 & 0.05 & 6 & 0 \\  
        BfS 41 & 70790962 & 270.488 & -30.022 & 05/09/2021 & 0.5 & 0.3 & 6.6 & 0.28 & 18.4 & 0 \\  
        BfS 42 & 70790975 & 270.723 & -29.525 & 14/11/2021 & 0.8 & 1.7 & 3.5 & 0.04 & 18.2 & 0 \\  
        BfS 43 & 70791011 & 270.369 & -29.511 & 15/08/2021 & 0.4 & 2.5 & 4.1 & 0.03 & 27.9 & 0 \\  
        BfS 44 & 70791023 & 270.899 & -29.293 & 15/05/2021 & 0.4 & 1.5 & 3.4 & 0.04 & 32.8 & 0 \\  
        BfS 45 & 70791355 & 272.603 & -36.434 & 31/08/2019 & 0.5 & 2.3 & 6.4 & 0.04 & 40.3 & 0 \\  
        BfS 46 & 70791434 & 272.517 & -37.488 & 16/08/2019 & 0.5 & 14.9 & 64.9 & 0.03 & 36.1 & 0 \\  
        BfS 47 & 70791686 & 273.023 & -36.568 & 23/08/2019 & 0.4 & 17.7 & 200.6 & 0.06 & 50.1 & 0 \\  
        BfS 48 & 70791761 & 271.248 & -37.247 & 31/08/2019 & 0.4 & 3 & 9.9 & 0.05 & 44.5 & 0 \\  
        BfS 49 & 70791852 & 272.860 & -37.199 & 31/08/2019 & 0.4 & 2.1 & 6.2 & 0.04 & 38 & 0 \\  
        BfS 50 & 70792117 & 272.207 & -37.403 & 31/08/2019 & 0.4 & 2.7 & 1.5 & 0.02 & 25.8 & 0 \\  
        BfS 51 & 70792231 & 237.067 & -48.184 & 31/07/2021 & 0.5 & 1 & 10.4 & 0.1 & 32.6 & 0 \\  
        BfS 52 & 70792375 & 235.628 & -47.398 & 17/10/2021 & 0.4 & 0.9 & 3.3 & 0.07 & 49.7 & 0 \\  
        BfS 53 & 70792445 & 236.614 & -47.322 & 31/10/2021 & 0.4 & 20.4 & 27.5 & 0.02 & 22.2 & 0 \\  
        BfS 54 & 70792463 & 236.839 & -47.273 & 17/10/2021 & 0.4 & 0.3 & 1.8 & 0.18 & 24.1 & 0 \\  
        BfS 55 & 70792507 & 236.701 & -47.932 & 19/06/2021 & 0.5 & 0.7 & 2.9 & 0.1 & 15.9 & 0 \\  
        BfS 56 & 70792542 & 235.884 & -47.378 & 17/10/2021 & 0.4 & 4.7 & 29.7 & 0.05 & 40.6 & 0 \\  
        BfS 57 & 70792578 & 236.787 & -47.672 & 27/09/2021 & 0.5 & 0.2 & 12.3 & 0.76 & 0.2 & 1 \\  
        BfS 58 & 70792640 & 236.045 & -47.552 & 17/10/2021 & 0.4 & 4.2 & 8.3 & 0.04 & 30.8 & 0 \\  
        BfS 59 & 70792656 & 236.271 & -47.244 & 07/08/2021 & 0.4 & 4.3 & 5.7 & 0.03 & 33.1 & 0 \\  
        BfS 60 & 70792673 & 237.774 & -47.166 & 31/07/2021 & 0.6 & 1.3 & 3.1 & 0.05 & 50.4 & 0 \\  
        BfS 61 & 70792680 & 237.816 & -47.732 & 31/07/2021 & 0.6 & 2.6 & 8.3 & 0.04 & 41.8 & 0 \\  
        BfS 62 & 70792689 & 236.974 & -48.085 & 23/10/2021 & 0.4 & 5.8 & 21.4 & 0.03 & 25.8 & 0 \\  
        BfS 63 & 70792709 & 236.908 & -47.147 & 23/10/2021 & 0.4 & 4.1 & 11.6 & 0.04 & 32 & 0 \\  
        BfS 64 & 70792753 & 237.413 & -47.324 & 31/10/2021 & 0.6 & 8 & 83 & 0.06 & 33 & 0 \\  
        BfS 65 & 70792803 & 237.946 & -47.945 & 31/07/2021 & 0.6 & 0.6 & 1.4 & 0.07 & 49.6 & 0 \\  
        BfS 66 & 70792836 & 237.318 & -48.068 & 19/06/2021 & 0.4 & 0.1 & 0.7 & 0.18 & 32 & 0 \\  
        BfS 67 & 70792856 & 235.627 & -47.468 & 17/10/2021 & 0.6 & 0.6 & 4.1 & 0.13 & 48.5 & 0 \\  
        BfS 68 & 70792874 & 236.501 & -47.879 & 31/07/2021 & 0.7 & 4 & 31.8 & 0.06 & 16.8 & 0 \\  
        BfS 69 & 70792923 & 235.831 & -47.516 & 05/09/2021 & 0.5 & 0.7 & 0.9 & 0.05 & 39.7 & 0 \\  
        BfS 70 & 70792950 & 237.244 & -48.086 & 23/10/2021 & 0.6 & 5.7 & 23.6 & 0.03 & 30.9 & 0 \\  
        BfS 71 & 70792982 & 237.261 & -47.217 & 31/10/2021 & 0.6 & 0.5 & 6.4 & 0.2 & 33.5 & 0 \\  
        BfS 72 & 70792990 & 237.207 & -47.712 & 31/07/2021 & 0.4 & 4 & 2.9 & 0.02 & 17.2 & 0 \\  
        BfS 73 & 70793015 & 235.650 & -47.575 & 04/10/2021 & 0.4 & 0.9 & 2.7 & 0.06 & 46.3 & 0 \\  
        BfS 74 & 70793035 & 235.482 & -47.815 & 31/10/2021 & 0.6 & 1.9 & 18.9 & 0.08 & 53.2 & 0 \\  
        BfS 75 & 70793087 & 282.321 & -3.065 & 07/09/2019 & 0.6 & 1 & 9528.9 & 1.53 & 0 & 1 \\  
        BfS 76 & 70793134 & 282.315 & -2.990 & 30/11/2019 & 0.4 & 74.9 & 11.7 & 0.02 & 4.5 & 0 \\  
        BfS 77 & 70793169 & 282.957 & -3.192 & 10/04/2020 & 0.4 & 5.9 & 8 & 0.04 & 38.9 & 0 \\  
        BfS 78 & 70793209 & 282.732 & -2.533 & 30/11/2019 & 0.5 & 6.7 & 7.6 & 0.03 & 40.3 & 0 \\  
        BfS 79 & 70793222 & 282.107 & -3.352 & 19/10/2019 & 0.4 & 37.6 & 8.1 & 0.01 & 21.5 & 0 \\  
        BfS 80 & 70793281 & 282.175 & -2.841 & 07/12/2019 & 0.4 & 3.6 & 54.5 & 0.32 & 16 & 0 \\  
        BfS 81 & 70793303 & 282.491 & -3.292 & 10/04/2020 & 0.4 & 0.8 & 3.2 & 0.21 & 17 & 0 \\  
        BfS 82 & 70793382 & 282.355 & -3.728 & 07/09/2019 & 0.4 & 13.8 & 9.9 & 0.03 & 39.8 & 0 \\  
        BfS 83 & 70793482 & 266.565 & -32.234 & 21/09/2018 & 0.6 & 0.4 & 146.3 & 1.09 & 0 & 1 \\  
        BfS 84 & 70793565 & 266.151 & -32.564 & 27/10/2018 & 0.4 & 0.6 & 1.5 & 0.22 & 28.9 & 0 \\  
        BfS 85 & 70793663 & 266.671 & -32.234 & 19/10/2018 & 0.4 & 0.2 & 26.8 & 1.18 & 5.4 & 1 \\  
        BfS 86 & 70793688 & 266.761 & -32.828 & 27/10/2018 & 0.5 & 0.8 & 1.6 & 0.12 & 37 & 0 \\  
        BfS 87 & 70793773 & 266.127 & -32.060 & 28/09/2018 & 0.4 & 1 & 6.7 & 0.23 & 24.6 & 0 \\  
        BfS 88 & 70793847 & 284.523 & -9.106 & 05/08/2019 & 0.5 & 2.1 & 8.9 & 0.05 & 52.6 & 0 \\  
        BfS 89 & 70793894 & 284.524 & -7.907 & 05/08/2019 & 0.6 & 3.3 & 6.4 & 0.03 & 21.1 & 0 \\  
        BfS 90 & 70793933 & 284.260 & -8.211 & 05/08/2019 & 0.7 & 2.4 & 8.1 & 0.04 & 22.9 & 0 \\  
        BfS 91 & 70794074 & 285.144 & -8.774 & 05/08/2019 & 0.6 & 0.8 & 9.2 & 0.11 & 43.7 & 0 \\  
        BfS 92 & 70794096 & 283.910 & -8.295 & 05/08/2019 & 0.4 & 0.9 & 4.2 & 0.08 & 43.8 & 0 \\  
        BfS 93 & 70794108 & 284.531 & -8.706 & 05/08/2019 & 0.7 & 3.1 & 14.9 & 0.04 & 28.9 & 0 \\  
        BfS 94 & 70794119 & 284.483 & -8.148 & 05/08/2019 & 0.4 & 5.6 & 16.9 & 0.02 & 11 & 0 \\  
        BfS 95 & 70794176 & 284.696 & -7.772 & 05/08/2019 & 0.4 & 0.6 & 1 & 0.07 & 28.1 & 0 \\  
        BfS 96 & 70794327 & 284.412 & -7.658 & 05/08/2019 & 0.6 & 0.4 & 1.1 & 0.08 & 37.4 & 0 \\  
        BfS 97 & 70794440 & 284.789 & -8.391 & 05/08/2019 & 0.4 & 0.3 & 0.7 & 0.11 & 12.5 & 0 \\  
        BfS 98 & 70794613 & 285.194 & -8.346 & 05/08/2019 & 0.5 & 1.5 & 5.5 & 0.05 & 33.2 & 0 \\  
        BfS 99 & 70794684 & 284.574 & -8.078 & 05/08/2019 & 0.7 & 3.3 & 30.7 & 0.05 & 10.5 & 0 \\  
        BfS 100 & 70794685 & 284.858 & -7.609 & 05/08/2019 & 0.7 & 1.1 & 55.5 & 0.23 & 39.8 & 0 \\  
        BfS 101 & 70794791 & 284.623 & -8.379 & 05/08/2019 & 0.9 & 2.3 & 21.8 & 0.05 & 8.6 & 0 \\  
        BfS 102 & 70794807 & 285.090 & -8.988 & 05/08/2019 & 0.8 & 0.5 & 5.1 & 0.14 & 52.1 & 0 \\  
        BfS 103 & 70794826 & 284.444 & -8.611 & 05/08/2019 & 0.9 & 1.8 & 12.5 & 0.06 & 25.4 & 0 \\  
        BfS 104 & 70794918 & 285.358 & -8.161 & 05/08/2019 & 0.6 & 1.2 & 6.6 & 0.07 & 42.5 & 0 \\  
        BfS 105 & 70794964 & 284.783 & -8.853 & 05/08/2019 & 0.4 & 1.1 & 2.3 & 0.05 & 37.8 & 0 \\  
        BfS 106 & 70795034 & 284.820 & -8.057 & 05/08/2019 & 0.8 & 1.1 & 6.7 & 0.07 & 15 & 0 \\  
        BfS 107 & 70795054 & 285.000 & -8.120 & 05/08/2019 & 1 & 2.3 & 21.6 & 0.07 & 22.2 & 0 \\  
        BfS 108 & 70795071 & 285.019 & -8.570 & 05/08/2019 & 0.4 & 0.5 & 2.9 & 0.12 & 29.8 & 0 \\  
        BfS 109 & 70795100 & 284.877 & -8.091 & 05/08/2019 & 0.7 & 0.7 & 5 & 0.13 & 16.3 & 0 \\  
        BfS 110 & 70795197 & 284.812 & -8.673 & 05/08/2019 & 0.8 & 1.9 & 8.4 & 0.06 & 27.9 & 0 \\  
        BfS 111 & 70795265 & 284.676 & -8.695 & 05/08/2019 & 0.4 & 1.8 & 2.1 & 0.03 & 27.5 & 0 \\  
        BfS 112 & 70795290 & 284.448 & -7.598 & 05/08/2019 & 0.4 & 2.9 & 14.3 & 0.05 & 40.1 & 0 \\  
        BfS 113 & 70795327 & 284.871 & -8.073 & 05/08/2019 & 0.8 & 9 & 103 & 0.05 & 16.6 & 0 \\
        BfS 114 & 70795334 & 284.800 & -7.904 & 05/08/2019 & 0.4 & 1.8 & 1.9 & 0.03 & 22 & 0 \\  
        BfS 115 & 70795365 & 284.708 & -8.128 & 05/08/2019 & 0.8 & 3.3 & 15.3 & 0.03 & 7.5 & 0 \\  
        BfS 116 & 70795426 & 284.341 & -7.904 & 05/08/2019 & 0.6 & 0.5 & 2.4 & 0.11 & 27 & 0 \\  
        BfS 117 & 70795476 & 284.619 & -7.956 & 05/08/2019 & 0.7 & 1.7 & 3.4 & 0.04 & 17 & 0 \\  
        BfS 118 & 70795913 & 207.308 & -62.922 & 17/11/2019 & 0.5 & 2 & 2.9 & 0.07 & 22.2 & 0 \\ 
        BfS 119 & 70795974 & 207.569 & -63.676 & 09/04/2019 & 0.4 & 20.9 & 50.9 & 0.04 & 27.8 & 0 \\  
        BfS 120 & 70796049 & 207.055 & -63.273 & 09/04/2019 & 0.7 & 0.2 & 434.9 & 1.72 & 0.1 & 1 \\  
        BfS 121 & 70796442 & 207.054 & -63.274 & 01/02/2019 & 0.4 & 0.5 & 2936.6 & 1.88 & 0 & 1 \\  
        BfS 122 & 70796444 & 207.993 & -63.692 & 01/03/2019 & 0.4 & 2.3 & 3.7 & 0.07 & 35.6 & 0 \\  
        BfS 123 & 70796614 & 274.548 & 7.188 & 14/10/2018 & 0.4 & 4.2 & 434.7 & 0.24 & 32.3 & 0 \\  
        BfS 124 & 70796677 & 275.168 & 7.804 & 14/10/2018 & 0.6 & 8 & 487.9 & 0.24 & 37.4 & 0 \\  
        BfS 125 & 70796699 & 274.869 & 6.912 & 14/10/2018 & 0.4 & 0.2 & 1.4 & 0.29 & 21.1 & 0 \\  
        BfS 126 & 70796701 & 274.509 & 7.811 & 14/10/2018 & 0.6 & 1.3 & 81.5 & 0.26 & 51.1 & 0 \\  
        BfS 127 & 70796797 & 275.120 & 6.698 & 14/10/2018 & 0.5 & 0.7 & 22.5 & 0.26 & 29.3 & 0 \\  
        BfS 128 & 70796831 & 274.919 & 6.673 & 14/10/2018 & 0.4 & 0.2 & 3.4 & 0.4 & 32.4 & 0 \\  
        BfS 129 & 70796842 & 274.773 & 7.007 & 14/10/2018 & 0.8 & 0.6 & 21.6 & 0.31 & 21.7 & 0 \\  
        BfS 130 & 70796845 & 274.883 & 7.940 & 14/10/2018 & 0.4 & 0.2 & 3.3 & 0.31 & 47 & 0 \\  
        BfS 131 & 70796863 & 274.588 & 7.085 & 14/10/2018 & 0.4 & 0.3 & 8.9 & 0.3 & 30.6 & 0 \\  
        BfS 132 & 70796900 & 275.511 & 7.878 & 14/10/2018 & 0.4 & 0.8 & 28.2 & 0.31 & 48.5 & 0 \\  
        BfS 133 & 70796970 & 274.727 & 7.523 & 14/10/2018 & 0.4 & 0.9 & 39.5 & 0.26 & 29.7 & 0 \\  
        BfS 134 & 70796978 & 275.166 & 7.056 & 14/10/2018 & 0.6 & 0.6 & 14.5 & 0.31 & 8.9 & 0 \\  
        BfS 135 & 70797098 & 275.110 & 6.915 & 14/10/2018 & 0.4 & 1.4 & 63.7 & 0.27 & 16.3 & 0 \\  
        BfS 136 & 70797121 & 275.105 & 7.526 & 14/10/2018 & 0.4 & 0.5 & 6.1 & 0.28 & 20.5 & 0 \\  
        BfS 137 & 70797142 & 275.235 & 6.442 & 14/10/2018 & 0.4 & 0.2 & 5.6 & 0.36 & 45.4 & 0 \\  
        BfS 138 & 70797182 & 274.707 & 6.479 & 14/10/2018 & 0.6 & 2.4 & 322.8 & 0.36 & 48.2 & 1 \\  
        BfS 139 & 70797198 & 275.717 & 7.629 & 14/10/2018 & 0.5 & 0.3 & 4.8 & 0.31 & 45.7 & 0 \\  
        BfS 140 & 70797292 & 274.609 & 6.771 & 14/10/2018 & 0.5 & 0.4 & 9.2 & 0.32 & 38 & 0 \\  
        BfS 141 & 70797295 & 274.819 & 6.766 & 14/10/2018 & 0.5 & 0.4 & 3.3 & 0.24 & 29.9 & 0 \\  
        BfS 142 & 70797361 & 275.038 & 7.447 & 13/11/2018 & 0.7 & 0.5 & 14.2 & 0.32 & 16 & 0 \\  
        BfS 143 & 70797376 & 274.875 & 7.784 & 14/10/2018 & 0.6 & 8.7 & 1052.6 & 0.26 & 38.2 & 0 \\  
        BfS 144 & 70797391 & 275.126 & 7.338 & 14/10/2018 & 0.4 & 0.4 & 6.3 & 0.3 & 9.4 & 0 \\  
        BfS 145 & 70797458 & 275.649 & 6.797 & 14/10/2018 & 0.7 & 0.8 & 8.2 & 0.32 & 40.5 & 0 \\  
        BfS 146 & 70797482 & 274.980 & 7.739 & 14/10/2018 & 0.7 & 0.7 & 36.4 & 0.37 & 33.8 & 0 \\  
        BfS 147 & 70797595 & 275.125 & 6.572 & 14/10/2018 & 0.7 & 3.8 & 520.9 & 0.28 & 36.9 & 1 \\  
        BfS 148 & 70797626 & 274.716 & 7.064 & 14/10/2018 & 0.6 & 0.1 & 2 & 0.32 & 23.5 & 0 \\  
        BfS 149 & 70797723 & 274.544 & 7.789 & 14/10/2018 & 0.4 & 0.3 & 8.7 & 0.28 & 48.7 & 0 \\  
        BfS 150 & 70797725 & 275.532 & 7.768 & 14/10/2018 & 0.5 & 1.1 & 35.8 & 0.26 & 43.7 & 0 \\  
        BfS 151 & 70797752 & 275.091 & 7.186 & 05/10/2018 & 0.7 & 0.2 & 1546.9 & 1.78 & 0 & 1 \\  
        BfS 152 & 70797802 & 275.526 & 6.549 & 14/10/2018 & 0.6 & 0.6 & 10.2 & 0.27 & 46.1 & 0 \\  
        BfS 153 & 70797809 & 275.818 & 7.101 & 14/10/2018 & 0.5 & 0.5 & 3.2 & 0.28 & 43.6 & 0 \\  
        BfS 154 & 70797837 & 274.955 & 7.512 & 14/10/2018 & 0.4 & 0.9 & 18.1 & 0.25 & 21.2 & 0 \\  
        BfS 155 & 70797847 & 275.930 & 7.263 & 14/10/2018 & 0.6 & 0.7 & 28.6 & 0.3 & 50.2 & 0 \\  
        BfS 156 & 70797867 & 274.860 & 7.798 & 14/10/2018 & 0.7 & 0.6 & 6.8 & 0.24 & 39.2 & 0 \\  
        BfS 157 & 70797932 & 275.014 & 7.750 & 14/10/2018 & 0.5 & 1.9 & 79.4 & 0.26 & 34.2 & 0 \\  
        BfS 158 & 70798003 & 275.227 & 7.283 & 14/10/2018 & 0.4 & 12 & 853.4 & 0.26 & 10 & 0 \\  
        BfS 159 & 70798081 & 275.063 & 7.475 & 14/10/2018 & 0.4 & 0.8 & 13.2 & 0.24 & 17.5 & 0 \\  
        BfS 160 & 70798095 & 274.520 & 6.958 & 14/10/2018 & 0.6 & 0.9 & 25.4 & 0.24 & 36.7 & 0 \\  
        BfS 161 & 70798104 & 274.469 & 6.777 & 10/04/2020 & 0.7 & 31.5 & 1686.3 & 0.3 & 44.4 & 1 \\  
        BfS 162 & 70798126 & 274.372 & 7.327 & 14/10/2018 & 0.4 & 0.4 & 13 & 0.3 & 43.6 & 0 \\  
        BfS 163 & 70798130 & 274.748 & 7.214 & 14/10/2018 & 0.6 & 1.3 & 81.3 & 0.27 & 20.5 & 0 \\  
        BfS 164 & 70798239 & 275.506 & 7.480 & 14/10/2018 & 0.5 & 0.3 & 4.4 & 0.31 & 30.4 & 0 \\  
        BfS 165 & 70798392 & 274.511 & 7.213 & 14/10/2018 & 0.5 & 1.1 & 96.5 & 0.3 & 34.6 & 0 \\  
        BfS 166 & 70798400 & 275.445 & 7.451 & 14/10/2018 & 0.5 & 0.1 & 1.3 & 0.38 & 26.4 & 0 \\  
        BfS 167 & 70798559 & 275.166 & 7.145 & 14/10/2018 & 0.6 & 8.9 & 409.4 & 0.24 & 5 & 0 \\  

\end{longtable}
\twocolumn


\bsp	
\label{lastpage}
\end{document}